\documentclass[review,12pt]{elsarticle}

\usepackage{lineno,hyperref}
\usepackage{amsmath}
\usepackage{graphicx}
\usepackage{caption}
\usepackage{mathtools}
\usepackage{esint}
\usepackage{upgreek}
\usepackage{float}
\restylefloat{table}
\usepackage{siunitx}
\usepackage{multirow}
\usepackage{bm}
\usepackage{xcolor}
\usepackage{siunitx}
\usepackage{lscape}
\usepackage{longtable}
\usepackage{gensymb}
\usepackage{setspace}
\usepackage{threeparttable}
\usepackage{tablefootnote}

\modulolinenumbers[5]

\biboptions{authoryear}

\begin{document}

\begin{frontmatter}

\title{Influence of Fluid Rheology on Fluid Flow in a Natural Fracture Network}

\author[Unimelb]{Cuong Mai Bui\corref{cor1}}
\author[NEU]{Stephan K. Matth{\"a}i\corref{cor2}}

\cortext[cor1,cor2]{Corresponding authors: 
	Cuong Mai Bui (cuong.bui@student.unimelb.edu.au), 
	Stephan K. Matth{\"a}i (stephan.matthai@icloud.com).}

\address[Unimelb]{Department of Infrastructure Engineering, The University of Melbourne,\\
	Grattan Street, Parkville, Victoria 3010, Australia}

\address[NEU]{School of Resource and Civil Engineering, Northeastern University,\\
	Shenyang 110819, China}

\begin{abstract}
	
Non-Newtonian rheology is widely acknowledged in subsurface fluids, yet its presence and effects are largely ignored in current fracture-flow studies. Here, we simulate fracture flow of non-Newtonian polymer solutions on a several metre-wide millimetre-aperture network of fractures, examining the complex interplay between fluid rheology, fracture geometry, and fluid inertia. Non-Newtonian fluid characteristics, including a yield stress and shear-thinning behaviour, are modelled using the Herschel-Bulkley-Papanastasiou approach. For the investigated flow rates, our Navier-Stokes simulations reveal significant viscosity variations, resulting in complex flow patterns at both aperture- and network scale. At low rates, non-yielded fluid forms rigid zones occupying up to $\sim 65\%$ of the network cross-sectional area, reducing fracture flow connectivity. At high rates, shear-thinning promotes inertia-dominated flow with circulations near fracture intersections. Regarding flow partitioning, yield stress confines flow to dominant pathways while shear-thinning promotes a broader fluid distribution as compared with a Newtonian fluid. Observed multimodal velocity distributions and nonlinear pressure drop-flow rate relationships underscore that fluid rheology must be considered during fracture-flow modelling.

\end{abstract}


\end{frontmatter}

\pagebreak


\section*{Highlights}

\begin{itemize}
	\item First simulations of non-Newtonian fluid flow in natural fracture networks.  
	\item Network geometry drives shear variations, causing viscosity contrasts. 
	\item Yield stress causes rigid blockages, enhancing flow localisation.  
	\item Shear-thinning increases network flow connectivity.  
	\item Fluid rheology fosters multimodal velocity distributions.  
\end{itemize}

\pagebreak

\section*{List of symbols and notations}

\begin{table*}[h]
	\begin{tabular}{p{1.5cm} p{7.5cm} p{2cm}}
		\hline
		\textbf{Symbol} & \textbf{Description} & \textbf{SI Unit} \\
		\hline
		$a$	& Fracture aperture & m \\
		$\mathrm{Bn}$ & Bingham number & - \\
		$d_f$ & Shear displacement & m \\
		$\Delta p$ &	Pressure drop	& Pa \\
		$J$ & Hydraulic head gradient & - \\
		$K$	 & Fluid consistency &	$\mathrm{Pa \cdot s^n}$ \\
		$L$ & Fracture length & m \\
		$m$ & Regularisation parameter &  - \\
		$\mu$ & Fluid dynamic viscosity & $\mathrm{Pa \cdot s}$ \\
		$\mu_{app}$ & Apparent viscosity & $\mathrm{Pa \cdot s}$ \\
		$\mu_c$ & Critical viscosity & $\mathrm{Pa \cdot s}$ \\
		$n$	 & Power-law index &	- \\
		$\mathrm{Re}$ &	Reynolds number &	- \\
		$\rho$ & Fluid density & $\mathrm{kg \cdot m^{-3}}$ \\
		$\tau$ & Shear stress & $\mathrm{Pa}$ \\
		$\tau_0$ & Yield stress & $\mathrm{Pa}$ \\
		$u_0$ &Injection rate & $\mathrm{m \cdot s^{-1}}$ \\
		$\bm{u}$ &	Volumetric flow rate & $\mathrm{m^3 \cdot s^{-1}}$ \\
		$\bm{v}$ & Velocity & $\mathrm{m \cdot s^{-1}}$ \\
		$w$ & 	Fracture width	& m \\
		$\dot{\gamma}$ & Shear rate & $\mathrm{s^{-1}}$ \\
		$\dot{\gamma_c}$ & Critical shear rate & $\mathrm{s^{-1}}$ \\
		\hline
	\end{tabular}
\end{table*}

\pagebreak

\section{Introduction}

Rock fractures form in response to a range of geological processes such as stresses induced by tectonics~\cite[][]{molnar2007tectonics}, thermal expansion and contraction~\cite[][]{richter1974thermal}, and weathering~\cite[][]{chigira2000mechanism}. Sequentially formed fracture sets coalesce to develop conduits for subsurface fluid flow~\cite[][]{parrish1963,berkowitz2002}. Each set has unique characteristics such as orientation, length-frequency distribution, spacing statistics and fracture attributes~\cite[][]{ramsay1980,nelson2001}. Sets abut or intersect, forming highly heterogeneous networks~\cite[][]{pollard1988,odling1997}. In hydraulic fracturing, opening-mode fractures are reactivated or created intentionally to enhance rock permeability and facilitate fluid transport~\cite[][]{hubbert1957mechanics,li2015review}. Due to the geometric complexity of fracture patterns, fluid flow within them exhibits substantial variations in velocity and flow regimes~\cite[][]{berkowitz2002,matthai2004,matthai2024}.


In reality, most fluids exhibit non-Newtonian rheology, i.e., a viscosity that varies with the rate of deformation~\cite[][]{di2010estimates,lavrov2023}. Common behaviours are shear-thinning (viscosity decreases with increasing shear rate), shear-thickening (viscosity increases with shear stress). Viscoplastic fluids require a minimum shear to flow. In nature, as is well-documented for crystal-rich magmas, the crystals arrange themselves during flow, depending on both shear rate and yield stress~\cite[][]{smith1997shear,smith2000textural,caricchi2007non,mader2013rheology,kolzenburg2022magma,vasseur2023shear}.

The oil and gas industry has shifted to using non-Newtonian fluids (e.g., polymers, foams, emulsions) for fracture stimulation~\cite[][]{barbati2016complex,osiptsov2017fluid,wrobel2021influence} to enhance oil recovery (EOR)~\cite[][]{wei2014oil,rellegadla2017polymers,firozjaii2020review}. Polymer flooding has become the most effective method of mobility control, improving sweep efficiency and potentially unlocking an additional $\sim 5-20\%$ of the original oil in place (OOIP) as compared to water flooding~\cite[][]{wang2009review}. Heavy oil also possesses non-Newtonian features (e.g., an extremely large yield stress of $\tau_0 \sim 150 - 200$~Pa~\cite{mendes2017yield}), complicating  flow and transport in fractured reservoirs~\cite[cf., ][]{coussot2014yield}.


Despite the widespread recognition of these non-Newtonian characteristics of reservoir fluids, fracture-flow studies have focused predominantly on Newtonian behaviour~\cite[][]{lavrov2023}. Amongst them, most experimental and numerical studies usually rely on simple fracture geometries~\citep[cf.,][]{hansika2024effects}, or even single-fracture setups~\cite[e.g., ][]{snow1970frequency,brown1998experimental,skjetne1999high,rong2020quantitative} or individual intersections~\cite[e.g., ][]{kosakowski1999flow,liu2018nonlinear,wu2018effects,xiong2020experimental}. This grossly underestimates geometric complexity. Some studies have considered fracture surface roughness~\cite[e.g., ][]{skjetne1999high,briggs2017numerical,kim2022flow,ma2022investigation,finenko2024numerical,finenko2024numericalb}, revealing irregular flow patterns controlled by variable aperture. Can roughness alone account for the flow structures seen in real-world scenarios? In complex fracture networks, where flow abruptly changes direction many times across variably-shaped intersections, the role of intersections may be far more significant~\cite[cf., ][]{liu2016b,zou2017modeling,matthai2024}. However, the limited studies on non-Newtonian fluids have only employed single-fracture models~\cite[e.g., ][]{yan2008flow,lavrov2013numerical,lavrov2013redirection,zhang20193d,zhang2023displacement}, restricting the analysis of flow physics to the cm-scale~\cite[][]{lavrov2023}. Focused narrowly on shear-thinning, these studies demonstrate several important effects such as enhanced flow localisation~\cite{lenci2022monte}, channeling~\cite[][]{lavrov2013redirection} and an accelerated transition to the nonlinear regime~\cite[][]{gao2024effect}, as compared to the Newtonian rheology.


While physical experiments are faced with many challenges due to the scale and complexity of relevant natural fracture patterns~\cite[e.g., ][]{renard2006characterization}, current fracture network-flow simulations rely on oversimplified models and modelling assumptions~\cite[][]{hansika2024effects}. A common simplification represents fractures as one-dimensional (1D) segments embedded in a 2D or 3D domain~\citep[e.g., ][]{lopes2022advancements,liem2022adaptive,tabrizinejadas2023coupled}. This reduces the problem to a graph-based flow model, significantly lowering the degrees of freedom associated with fracture voids and hence enabling large-scale simulations~\citep[][]{matthai2004}. However, this simplification preempts the study of fracture internal flow structures and affects physical accuracy. Geometrically, the reduced-order representation of fractures also neglects fracture curvature related centrifugal forces, and surface roughness~\cite[cf., ][]{hansika2024effects}. Intersections among fractures are reduced to point or line manifolds, ignoring entry angles, offset junctions, and intersection areas that control flow redistribution. As a result, realistic fracture-flow patterns such as flow separation, recirculation zones, and vortex formation cannot be captured~\citep[cf., ][]{kim2022flow,cardenas2009effects}. Inertia effects, including transient dynamics and local under-pressure phenomena, critical for predicting energy losses and transport heterogeneity~\citep[cf., ][]{skjetne1999high,matthai2024,bui2025dynamics}, are ignored entirely.


The single-fracture-derived cubic law for laminar creeping flow of a Newtonian fluid~\cite[][]{zimmerman1996hydraulic,zimmerman2000fluid,konzuk2004evaluation} is widely applied to estimate network pressure drops relying on its linear relationship with volumetric flow rate~\cite[e.g., ][]{ge1997governing,min2004determining,noetinger2015quasi}. Despite many efforts to extend the range of applicability of this law, primarily by accounting for additional losses from fracture roughness~\cite[][]{neuzil1981flow,brown1987fluid,wang2016influence} and fluid inertia such as in the full cubic law~\cite[e.g., ][]{fourar1993two,nowamooz2009non} or by using Forchheimer's formula~\cite[e.g., ][]{wu2018effects,fan2019effects}, the nonlinearity associated with fluid rheology remains unaddressed~\cite[cf., ][]{tosco2013extension,roustaei2016non}. The applicability to fracture networks of a modified version of Darcy's law, developed to account for non-Newtonian flow effects in porous media~\cite[e.g., ][]{mckinley1966non,pearson2002models,tsakiroglou2002methodology}, is highly questionable, given their orders-of-magnitude higher permeability~\cite[][]{adler2013fractured}. \cite{matthai2024} and \cite{bui2025dynamics} presented evidence that such macroscopic approaches fail to capture aperture-scale flow behaviours and their influence on flow velocity distributions within fracture networks.


Numerical fracture-network flow studies mostly approximate the Stokes' lubrication equation, addressing only the fluid diffusion term~\cite[e.g., ][]{koudina1998permeability,bogdanov2003effective,matthai2004,xu2021modelling}, and disregard the influence of fluid inertia, fluid rheology, and their interplay. Navier-Stokes simulations, considering fluid convection, in fracture networks are few, primarily due to the challenges of discretising complex geometries~\cite[][]{reiter2012preparation} and solving nonlinear partial differential terms in these patterns~\cite[][]{yu2017review,berre2019flow}. These difficulties are further exacerbated when dealing with non-Newtonian fluids~\cite[][]{lavrov2023} because nonlinear constitutive relationships defining the shear stress tensor introduce additional complexities to the governing equations~\cite[][]{chhabra2010non}. Recently, \cite{matthai2024} conducted Navier-Stokes-based simulations of water flow within natural fractures, attempting to take fluid inertia effects into account. This study and that of~\cite[][]{bui2025dynamics} revealed that different flow regimes coexist in a network, ranging from laminar creeping flow in stagnant fractures to turbulent flow. Frequent interactions between flow and fracture irregularities occur. 


To date, non-Newtonian inertia-affected fluid flow within realistic fracture patterns has yet to be investigated, leaving a critical gap in understanding how fluid rheology influences fracture flow dynamics. In a complex fracture system, flow velocity distribution is multimodal~\cite[][]{matthai2004,matthai2024}, spanning many orders of magnitude. Does such significant velocity variation and associated variation in shear rates imply that the influence of non-Newtonian rheology becomes more pronounced at the network scale, impacting fracture flow behaviour and distribution? Additionally, non-Newtonian fracture-flow studies often consider power-law shear-dependent viscosity only~\cite[][]{lavrov2023}, neglecting the yield stress inherent in engineering fluids such as polymers~\cite[][]{song2006rheology,jang2015enhanced,ghoumrassi2016rheological,sarmah2022formulation} or foams~\cite[][]{reidenbach1986rheological,kam2002yield,edrisi2012new,xu2020flow}. These studies are limited to low flow rates, not relevant for subsurface engineering activities such as polymer flooding or hydraulic fracturing, where high flow velocities are inducing fluid inertia effects~\cite[][]{chang1978polymer,edirisinghe2024review,edirisinghe2024particle}.

Our study aims to overcome this research gap by studying non-Newtonian fluid flow within a natural fracture pattern. The fracture geometry studied was mapped by~\cite{odling1997} in the Devonian Hornelen basin, Norway. This network features significant variability, including rich intersection configurations and dead-ends, lending itself to capturing key characteristics of realistic fracture flow. Aqueous polymer solution engineered for enhanced oil recovery, specifically Xanthan gum solutions with their shear-thinning and yield stress characterised in rheometer experiments by~\cite{jang2015enhanced}, are utilised. Navier-Stokes flow simulations with these agents are setup to span a wide range of influx rates, enabling a detailed investigation of fracture-network flow, considering fluid rheology, fluid inertia, network complexity, and the interplay among them. Contrasting and comparing with Newtonian fluids and among non-Newtonian polymer solutions, the influence of non-Newtonian rheology on flow behaviour, velocity distribution, and network flow partitioning, is examined and analysed.

We begin by describing the conceptual model, numerical approach, and simulation setups for network flow of non-Newtonian polymer solutions (Section~\ref{sec:Method}). Following a mesh sensitivity analysis and simulator verification, Section~\ref{sec:Section3} presents simulation results from both single-intersection and large-scale fracture network models, evaluating the influence of fluid rheology for different flow scenarios. A discussion of novel findings and concluding remarks are given in Sections~\ref{sec:Section4} and \ref{sec:Section5}, respectively.






\section{\label{sec:Method}Methodology}

\subsection{Governing and constitutive equations}




Incompressible fluid flow within fractures is governed by the continuity equation, representing mass conservation:

\begin{equation}
	\nabla \cdot \bm{v} = 0 \, .
	\label{eq_continuity}
\end{equation}

The momentum equation expresses the force balances between inertial (second term), pressure (third), and viscous forces (fourth):

\begin{equation}
	\rho \left( \dfrac{\partial \bm{v}}{\partial t} + \bm{v} \cdot \nabla \bm{v} \right) = - \nabla p +  \nabla \cdot \bm{\tau} \, .
	\label{eq_momentum} 
\end{equation}

Here, $\bm{v}$ is the velocity vector ($\mathrm{m \cdot s^{-1}}$), $\rho$ the fluid density ($\mathrm{kg \cdot m^{-3}}$), $t$ time ($s$), and $p$ fluid pressure (Pa), respectively. To simulate flow over a wide range of velocities, Equations~\ref{eq_continuity} and \ref{eq_momentum} are solved numerically using the Reynolds-averaged Navier-Stokes approach (RANS). RANS expresses inertia-driven flow without requiring highly refined meshes. This method is considered suitable for complex geometries such as fracture networks, where local Reynolds numbers can reach up to $10^4$, indicating the occurrence of strong turbulence~\cite[][]{matthai2024}. Further details of the RANS formulation and its implementation are provided in our previous studies~\cite[][]{matthai2024,bui2025influence}, among these the verification for a Newtonian aqueous fluid.

The constitutive equation for a Newtonian fluid defines the shear stress tensor as:

\begin{equation}
	\bm{\tau} = \mu \dot{\bm{\gamma}} \, ,
	\label{eq_tau}
\end{equation}

where $\mu = \mathrm{const}$ is the fluid viscosity ($\mathrm{Pa \cdot s}$) and $\dot{\bm{\gamma}}$ the strain rate tensor. For a non-Newtonian fluid with yield stress and shear-dependent behaviour, $\bm{\tau}$ can be estimated using the Herschel-Bulkley (HB) model as~\cite[]{herschel1926}:

\begin{equation}
	\begin{cases}
		\bm{\tau} = \left(K {\dot{\gamma}}^{n-1} + \dfrac{\tau_0}{\dot{\gamma}} \right) \dot{\bm{\gamma}} & \text{if } \tau > \tau_0 \\
		\dot{\gamma} = 0 & \text{if } \tau \leq \tau_0 \, .
		\label{eqn_HB}
	\end{cases}
\end{equation}

Here, $K$ models the fluid consistency ($\mathrm{Pa \cdot s^n}$), the power-law index $n$ controls the curvature of the $\dot{\gamma}$-$\tau$ relationship, and $\tau_0$ the yield stress threshold ($\mathrm{Pa}$). The term in bracket represents the apparent viscosity $\mu_{app}$, comprising the shear-rate-dependent power-law component and the yield-stress contribution. The strain rate ($\mathrm{s^{-1}}$) and shear stress ($\mathrm{Pa}$) magnitudes are computed as $\dot{\gamma}=\sqrt{\frac{1}{2} \dot{\bm{\gamma}}:\dot{\bm{\gamma}}}$ and $\tau=\sqrt{\frac{1}{2} \dot{\bm{\tau}}:\dot{\bm{\tau}}}$, respectively. For $\tau \leq \tau_0$, the fluid remains unyielded, exhibiting solid-like behaviour and resisting deformation ($\dot{\gamma}=0$) \cite[cf., ][]{alexandrou2001}. When $\tau > \tau_0$, the fluid yields, allowing continuous deformation and initiating flow. It then shears according to a power-law relationship. A limitation of this assumption is its inability to capture viscosity plateaus at very low and very high shear rates~\citep[cf., ][]{shahsavari2015mobility,boyko2021flow}. More sophisticated models, such as Ellis~\citep[][]{bird19770} or Carreau–Yasuda~\citep[][]{bird1987dynamics}, can represent these limiting viscosities but require additional experimentally determined parameters~\citep[][]{raju1993assessment,irgens2014rheology}. In this study, the rheometer data for EOR polymer aqueous solutions were well matched by the HB model~\cite{jang2015enhanced}, but the extreme low- and high-shear regimes were not characterised. For the low-shear conditions relevant to this work, the yield-stress behaviour is more critical than the detailed shear-rate dependence. Therefore, the thus-extended HB model is adequate for the present analysis.

To smooth the $\dot{\gamma}$-$\tau$ curve, avoiding numerical artifacts from the discontinuity at $\tau=\tau_0$ in the HB model~\cite[]{beverly1989}, the Papanastasiou's regularisation scheme is adopted~\cite[]{papanastasiou1987}:


\begin{equation} 
	\bm{\tau} = \left( K {\dot{\gamma}}^{n-1} + \frac{\tau_0 [ 1 - \exp{(-m \dot{\gamma})} ]}{\dot {\gamma}} \right) \bm{\dot{\gamma}} \, ,
	\label{eq_papa}
\end{equation}

with $m$ the regularisation parameter, selected to ensure a non-zero shear rate ($\dot{\gamma} = \dot{\gamma_c}$ with $\dot{\gamma_c}$ the critical shear rate) at the yield point ($\tau =\tau_0$), providing a realistic representation of yielding~\cite[]{bui2019}. The unyielded regions are thus defined as $\tau \leq \tau_0$~\cite[]{burgos1999} or $\dot{\gamma} \leq \dot{\gamma_c}$~\cite[]{bui2019,bui2021effects}.

Dimensionless numbers characterising flow in a single fracture including Reynolds number, represent the ratio of inertial to viscous effects~\citep[][]{mossaz2012experimental}:

\begin{equation}
	\mathrm{Re}= \dfrac{\rho {u_0}^{2-n} d^n}{K} \, ,
	\label{eqn_Re}
\end{equation}

where $u_0$ is the influx into the model, and $d$ a characteristic length, chosen as the fracture aperture $a$. For fluids exhibiting a yield stress, the Oldroyd number quantifies the relative importance of yield stress compared to viscous effects as~\citep[][]{mossaz2012experimental}:

\begin{equation}
	\mathrm{Od}= \dfrac{\tau_0 d^n}{K {u_0}^n} \, .
	\label{eqn_Od}
\end{equation}

In the case of a Bingham fluid, where only yield stress is present without additional shear-rate dependence ($n=1$), the Oldroyd number reduces to the Bingham number~\cite[]{jossic2002segregation}:

\begin{equation}
	\mathrm{Bn} = \dfrac{\tau_0 d}{K u_0} \, .
\end{equation}

These dimensionless numbers are applied solely to an idealised single intersection. For fracture-network flow, we do not use these dimensionless numbers to generalise flow behaviour across the network, as flow patterns are highly geometry-specific and strongly influenced by connectivity, aperture variability, and intersection structure~\citep[][]{matthai2024,bui2025dynamics}.

Additionally, the hydraulic head gradient, non-dimensionalising the pressure drop with gravitational force, is given by:

\begin{equation}
	J = \dfrac{\nabla p}{ \rho g} \, ,
\end{equation}

where $g = 9.81~\mathrm{m~s^{-1}}$ is the acceleration due to gravity.

\subsection{Fluid properties}

\begin{figure} 
	\centering
	\includegraphics[width=0.6\columnwidth]{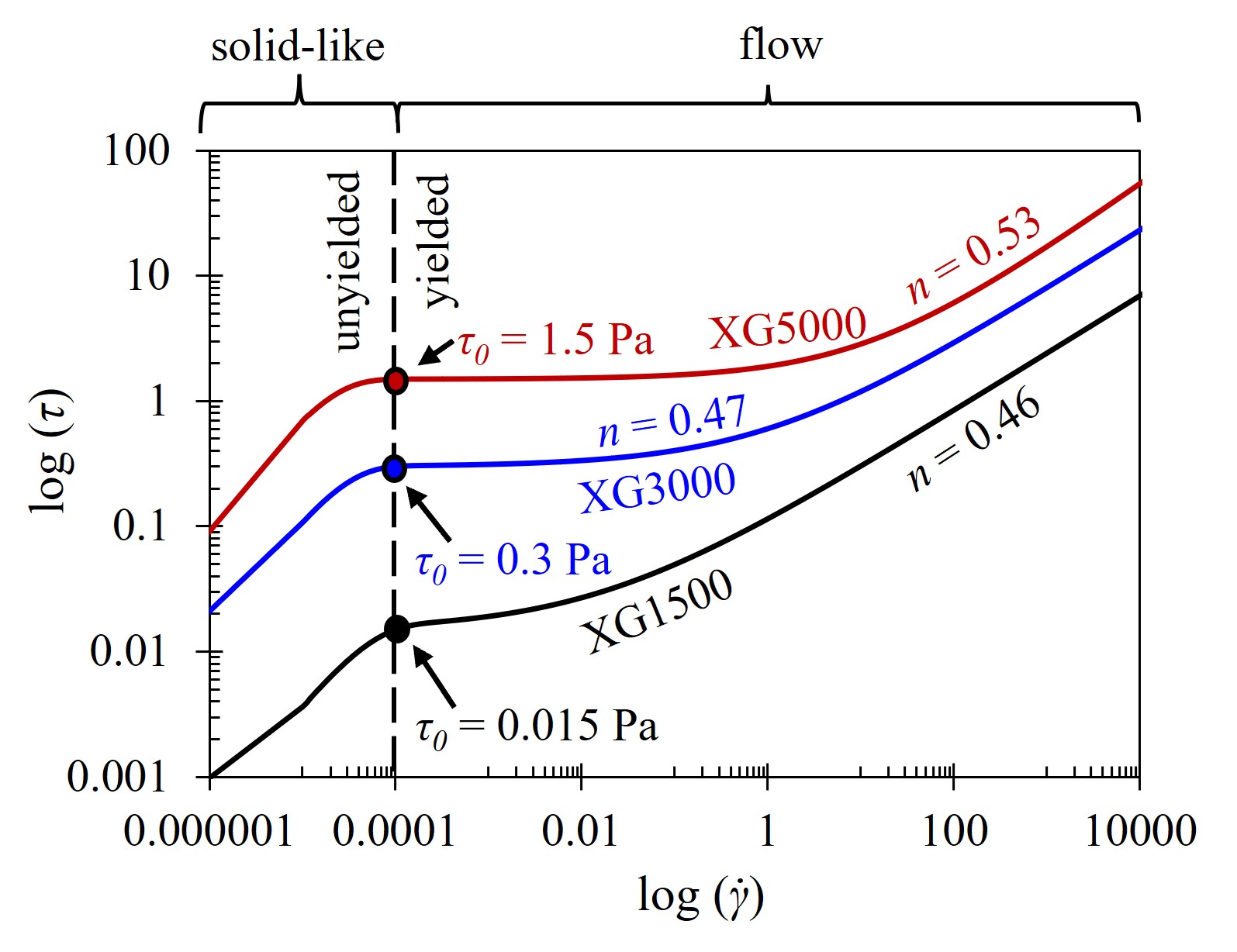}
	\caption{Rheological curves of xanthan gum solutions modelled using the Herschel-Bulkley-Papanastasiou approach. At low shear rates ($\dot{\gamma_c} \leq 0.0001~\mathrm{s^{-1}}$), yield stress dominates rheological response, making the fluid unyielded. At high shear-rates, yielded fluid follows the power-law shear-dependence.}
	\label{fig:Rheo_curve}
\end{figure}

Real-world non-Newtonian fluids inspiring our flow simulations include aqueous xanthan gum solutions, produced by cellulose fermentation with the bacterium \textit{Xanthomonas campestris} and widely used as an EOR agent as well as in hydraulic fracturing~\cite[][]{barbati2016}. Their rheological properties at 20\degree C and 3wt\% salinity, and varying xanthan concentrations, were measured and reported by~\cite{jang2015enhanced}. As mentioned before, rheological data of these solutions were perfectly fitted with the HB model; parameters for each concentration (i.e., $c = 1500, 3000,$ and 5000~ppm) are provided in Table~\ref{tb:XG}. Since the critical shear rate of xanthan polymers is not available, we assume it to be the same as that of the synthetic polymer solution Carbopol, with $\dot{\gamma_c} = 0.0001~\mathrm{s^{-1}}$, as measured by~\cite{mossaz2012experimental}. In this work, the focus is on capturing the rheological response and flow characteristics of industrial fluids within a complex network geometry rather than performing a broad parametric study; therefore, the ranges of $\tau_0$ and $n$ are not extensively varied, but reflect realistic value ranges, see Fig.~\ref{fig:Rheo_curve}.

Additionally, entirely hypothetical fluids are considered: (1) Newtonian fluids ($\tau_0 = 0$, $n=1$), (2) Power-law fluids ($\tau_0$) with varying $n$ values, and (3) Bingham fluids ($n=1$) with varying $\tau_0$; all using the same $K$ as the real xanthan solutions. These additional fluids are used for model verification and to isolate non-Newtonian effects in the simplified, single-intersection model, enabling to examine each fluid feature independently. Using these artificial fluids, shear-thinning and yield-stress properties are first analysed separately before assessing their combined effects on flow in complex fracture geometries. For labelling, Newtonian fluids are denoted with the suffix 'N' (e.g., XG5000N), power-law fluids with 'P', and Bingham fluids with 'B'. All fluids and their specific purposes in this study are summarised in Table~\ref{table:fluids}.

\begin{table}
	\centering
	\begin{tabular}{lcccc}
		\hline
		\textbf{Fluid} & $c$ ($\mathrm{ppm}$) & $K$ ($\mathrm{Pa \cdot s^n}$) & $\tau_0$ (Pa) & $n$ \\
		\hline
		XG1500 & 1500 & 0.1 & 0.015 & 0.46 \\
		XG3000 & 3000 & 0.3 & 0.3 & 0.47 \\
		XG5000 & 5000 & 0.401 & 1.5 & 0.53 \\
		\hline
	\end{tabular}
	\caption{\label{tb:XG} Herschel-Bulkley parameters for xanthan gum solutions at various concentrations as reported by ~\cite{jang2015enhanced}.}
\end{table}

\begin{table}
	\begin{tabular}{lllccc}
		\hline
		\textbf{Fluid} & \textbf{Description} & \textbf{Task} & \textit{K} \text{(Pa$\cdot$s$^n$)} & \textit{$\tau_0$} \text{(Pa)} & \textit{n} \\
		\hline
		XG1500 & HB & a & 0.1 & 0.015 & 0.46 \\
		XG1500B & Bingham & b & 0.1 & 0.015 & 1 \\
		XG1500N & Newtonian & a, b & 0.1 & 0 & 1 \\
		XG1500P & Power-law & b & 0.1 & 0 & 0.46 \\
		XG3000 & HB & a & 0.3 & 0.3 & 0.47 \\
		XG3000B & Bingham & b & 0.3 & 0.3 & 1 \\
		XG3000N & Newtonian & a & 0.3 & 0 & 1 \\
		XG5000 & HB & a & 0.401 & 1.5 & 0.53 \\
		XG5000B & Bingham & b, c & 0.401 & $0.1-1.5$ & 1 \\
		XG5000N & Newtonian & a, c & 0.401 & 0 & 1 \\
		XG5000P & Power-law & b, c & 0.401 & 0 & $0.4-0.8$ \\
		\hline
	\end{tabular}
	\caption{\label{table:fluids} Fluids used in this work and their properties.}
	\begin{tablenotes}
		\item a: Network flow investigation
		\item b: Model verification
		\item c: Single-intersection analysis
	\end{tablenotes}
\end{table}

\subsection{Discrete fracture network model}

\begin{figure} 
	\centering
	\includegraphics[width=0.75\columnwidth]{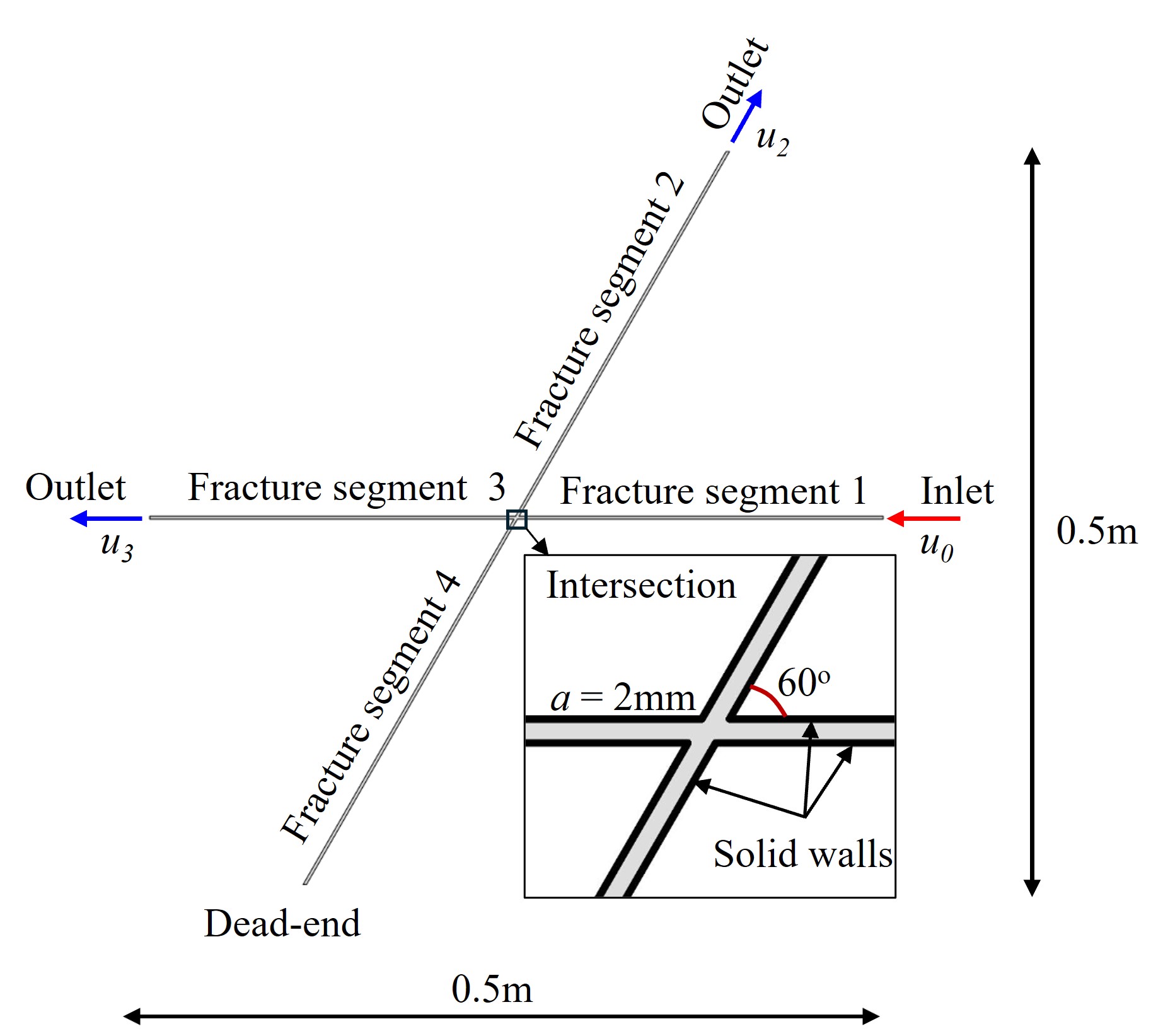}
	\caption{Geometry of the one-inlet, two-outlet X-shaped intersection model. The fracture aperture is 2~$\mathrm{mm}$, with an intersecting angle of $60\degree$. For verification, fracture 2 is closed, reducing the model to a one-inlet, one-outlet channel.}
	\label{fig:Xmodel_geometry}
\end{figure}

One X-shaped single-intersection model, featuring a branching angle of $60\degree$, serves the purpose of model verification and independent analysis of non-Newtonian effects. The model consists of smooth, parallel fracture surfaces, with a total length of 0.5~m and a constant separation (aperture) of 2~mm (Fig.~\ref{fig:Xmodel_geometry}). For the analysis of non-Newtonian fluid flow, the model is configured as a single-inlet, double-outlet geometry, with one branch (fracture 4) closed to create a dead-end. For the verification runs, both winding branches (fracture segments 2 and 4) are closed, further simplifying the geometry to resemble a single-channel fracture model.

\begin{figure} 
	\centering
	\includegraphics[width=\columnwidth]{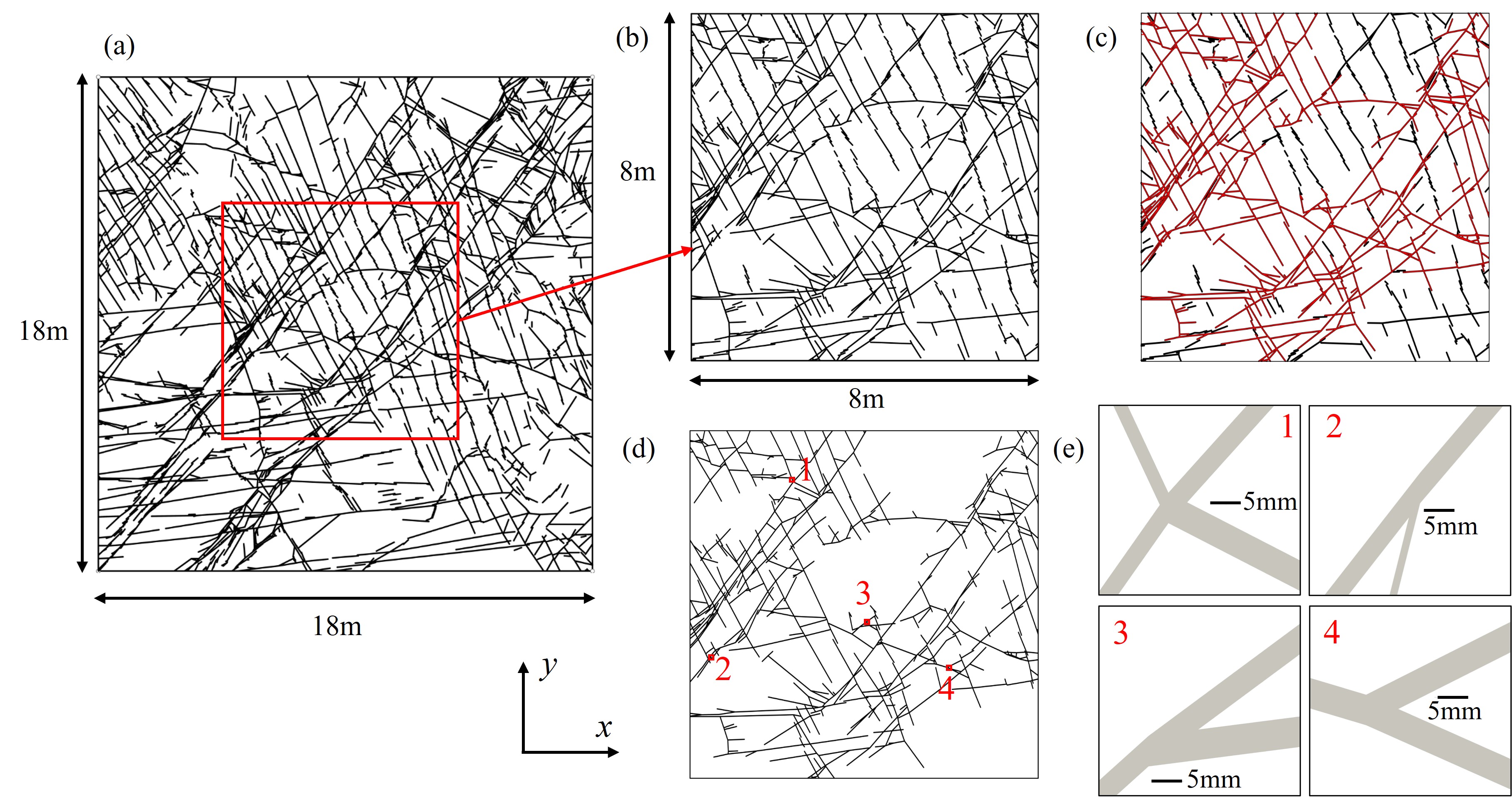}
	\caption{(a) Birds-eye view of the natural fracture network in the Devonian basin of Hornelen, western Norway~\citep[][]{odling1997}. (b) $8 \times 8~\mathrm{m}$ subregion to which a dextral shear displacement of 5 mm was applied, opening fracture void spaces, creating flow conduits. (c) Interconnected fractures are highlighted in red. (d) Discrete fracture network model constructed from the interconnected fractures shown in (c). (e) Zoomed-in view of fracture intersections at positions 1-4 in (d). }
	\label{fig:OH_geometry}
\end{figure}


To capture potentially multimodal flow velocity distribution and diverse flow regimes~\cite[cf., ][]{matthai2004,matthai2024}, we employ a field-derived discrete fracture network (DFN) model with complex geometrical features, including irregular intersections, dead-ends, and varying apertures. The natural $8 \times 8$~m fracture pattern stems from the $18 \times 18$~m outcrop line-tracing of~\cite{odling1997} in the Devonian Hornelen basin, western Norway (Fig.~\ref{fig:OH_geometry}). This network comprises three dominant fracture sets oriented approximately $0^\circ$, $50^\circ$, and $130^\circ$\cite[][]{odling1997}. Together these give rise to the characteristic structural anisotropy of the outcrop pattern. The network is geometrically highly connected, with numerous intersections linking individual fractures into a single dominant cluster. More than 90\% of the total fracture trace length belongs to this cluster, indicating complete percolation within the domain. Fracture connectivity is further enhanced by a dense population of short fractures bridging larger ones, producing a well-integrated cluster structure that supports continuous pathways for both flow and stress transmission.

To create void spaces, a shear displacement of $d_f = 5~\mathrm{mm}$ was applied to all fracture segments, extruding lines to a planar representation which accommodates fluid flow while retaining the original geometric characteristics. Since the fractured rock blocks now are essentially impermeable, disconnected fractures, which do not contribute to the network flow, are excluded from the flow model (Fig.~\ref{fig:OH_geometry}b, c). The resulting simulation model, consists of interconnected variable-aperture fractures channeling the flow, see Figure~\ref{fig:OH_geometry}d. In total, our network model contains 431 intersections, comprising 281 ($\sim 65.2\%$) Y- or T-shaped intersections, 139 ($\sim 32.3\%$) X-shaped intersections, and 11 of other irregular types ($\sim 2.5\%$).

Rock surfaces are idealised as smooth-walled to circumvent issues related to fracture roughness in a large-scale two-dimensional (2D) flow geometry. This simplification is justified, as the influence of small-scale roughness becomes negligible in complex fracture systems~\citep{liu2016b}. Fracture intersections themselves introduce abrupt geometric transitions that act as macroscopic roughness elements, exerting a dominant influence on flow resistance and mixing~\citep{hansika2024effects}. Furthermore, this model focuses on planar flow, assuming that fracture depth has a minimal impact~\cite[cf., ][]{liu2016b} on the flow, while gravitational effects are neglected. 


\subsection{Model discretisation and computational implementation}

\begin{figure} 
	\centering
	\includegraphics[width=\columnwidth]{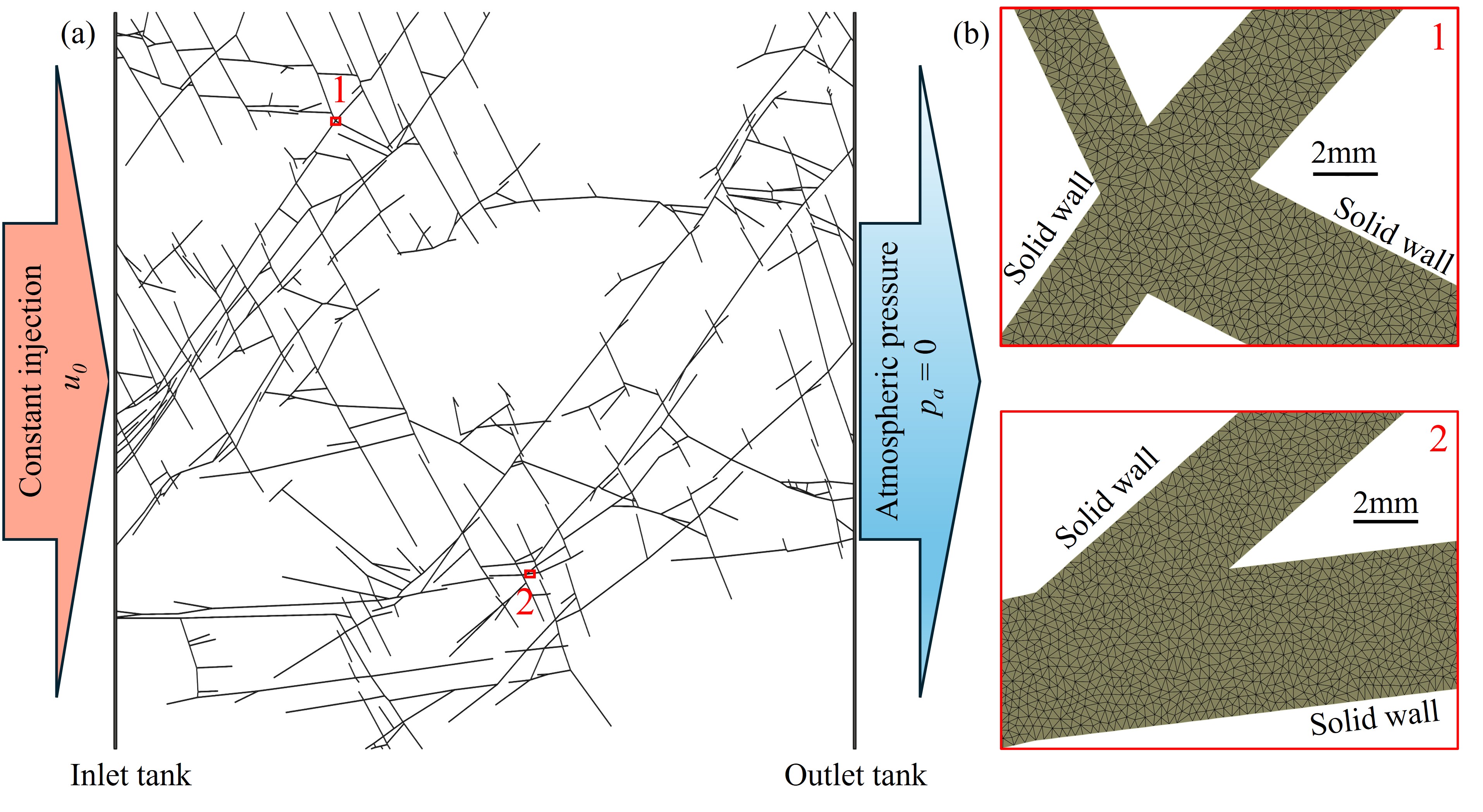}
	\caption{(a) Simulation setup for finite-volume (FV)-based flow simulations in the Hornelen network model. Inlet and outlet reservoirs are placed at the inflow and outflow surfaces to distribute fluid into and out of the network. (b) A closer view of the mesh at typical fracture intersections. The mesh consists of a total of 15M finite elements.}
	\label{fig:OH_mesh}
\end{figure}

As in our previous work~\cite[][]{matthai2024}, we discretise fracture geometries with triangular finite elements. Using Ansys ICEM CFD, for the X-model, a mesh convergence study is carried out, comparing simulation results obtained for three mesh resolutions with maximum element sizes of $a/20$, $a/25$, and $a/32$ where $a$ refers to fracture aperture. Results are presented in Section~\ref{Result:Veri}. For the Hornelen network model, we used the Triangle mesher, developed by~\cite{shewchuk1996b}, enabling fracture-only meshing. This approach conserves memory resources by restricting meshing to the millimetre-wide void spaces within the fractures where flow occurs, while excluding the impermeable rock matrix~\cite[][]{matthai2024}. The boundary conditions used in our simulation are illustrated in Fig.~\ref{fig:OH_mesh}. To allow a natural fluid distribution among fractures, inlet and outlet tanks are added at the network's entry and exit. Constant influxes, stepped from $u_0 = 10^{-5}$ to $10^{-2}~\mathrm{m^3 m^{-2} s^{-1}}$, are used to examine the effects of non-Newtonian fluid behaviour across different flow scenarios. The atmospheric pressure is prescribed at the outlet, creating a left-to-right far-field pressure gradient. Fracture walls are modelled with a stationary, impermeable boundary condition, confining flow to within the fracture channels.

We use a steady-state solver with a coupled pressure-velocity algorithm for the NS approximation. Second-order accuracy schemes are employed for fluid velocity and pressure. Finite volume computations are performed using the commercial simulation tool Ansys Fluent 2023R1. A user-defined function (UDF) has been developed to implement the Herschel-Bulkley-Papanastasiou constitutive relation (Eq.~\ref{eq_papa}), computing the apparent viscosity from the local strain-rate invariant with the regularisation parameter set at the yield point.

\subsection{Verification approach}

For model verification, simulated channel-flow velocity profiles of non-Newtonian fluids are compared with available analytical solutions at various flow velocities. For Bingham fluids, the analytical solution, derived by~\cite{bird1983}, is:

\begin{equation}
	\begin{cases} 
		v(y, \tau_0) = \dfrac{\Delta p B^2}{ 2 K L} \left[ 1 - \left( \dfrac{y}{B} \right)^2 \right] - \dfrac{\tau_0 B}{K} \left[ 1 - \left( \dfrac{y}{B} \right) \right] & \text{if } y_0 \leq y \leq B \\
		v(y, \tau_0) = \dfrac{\Delta p B^2}{ 2 K L} \left[ 1 - \left( \dfrac{y_0}{B} \right)^2 \right] - \dfrac{\tau_0 B}{K} \left[ 1 - \left( \dfrac{y_0}{B} \right) \right] & \text{if } 0 \leq y \leq y_0	\, .
	\end{cases}
	\label{eqn_Bingham_sol}
\end{equation}

where $\Delta p$ is the pressure drop along the fracture channel, $B =  a/2$ with $a$ the fracture aperture, $L$ the fracture length, and $y_0 = \tau_0 L / \Delta p$. For power-law fluids, the analytical velocity profile provided by~\cite{tanner2000} is used:

\begin{equation}
	v(y,n) = \left( \dfrac{2n+1}{n+1} \right) \left[ 1 - \left( \dfrac{y}{B} \right)^{1 +1/n}  \right] u_0 \, .
	\label{eqn_pl_sol}
\end{equation}

Notably, when $\tau_0 = 0$ in Eq.~\ref{eqn_Bingham_sol}, and when $n=1$ in Eq.~\ref{eqn_pl_sol}, the solution reduces to the fully-developed parabolic profile of a Newtonian fluid flow.

\section{\label{sec:Section3}Results}

\subsection{\label{Result:Veri}Mesh convergence and verification results}

\begin{figure} 
	\centering
	\includegraphics[width=0.9\columnwidth]{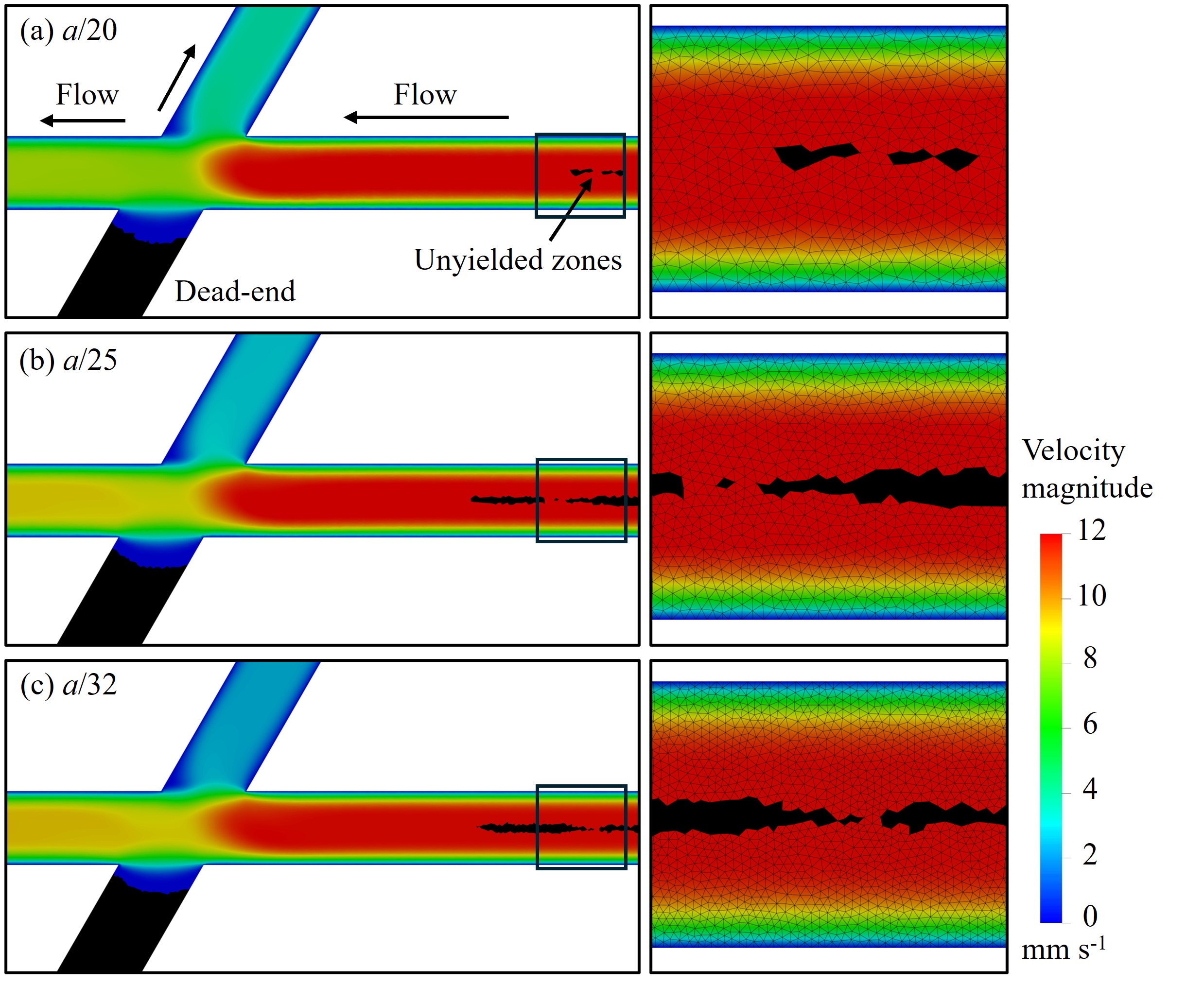}
	\caption{X-model: Variations in flow pattern of fluid XG5000 with different meshes at $u_0 = 1~\mathrm{cm~s^{-1}}$ ($\mathrm{Re} \sim 0.11$ and $\mathrm{Od} \sim 1.59$). The low mesh resolution in (a) underestimates the extent of unyielded zones ($\tau \leq \tau_0$).}
	\label{fig:Xmodel_meshconv_uz}
\end{figure}

The mesh sensitivity analysis with the X-model compared flow morphology, maximum velocity, and total pressure drop for different meshes. At all mesh resolutions, various irregular structures appeared within the fracture channel, corresponding to unyielded regions where $\tau \leq \tau_0$\cite[][]{burgos1999} or $\dot{\gamma} \leq \dot{\gamma_c}$~\citep[][]{bui2019} (Fig.~\ref{fig:Xmodel_meshconv_uz}). However, their size varied. At $a/20$, these regions were rather small (Fig.~\ref{fig:Xmodel_meshconv_uz}a) while for $a/25$ and $a/32$, a relatively good agreement was obtained, suggesting that refinement was sufficient to delineate them accurately. This assessment is corroborated by the quantitative results, where the maximum deviation in pressure drop relative to $a/32$ was $\sim$1.1\% for $a/20$ but only $\sim$0.2\% for $a/25$ (Table~\ref{tab:mesh_conv}). Given the significantly larger computational time required for $a/32$, $a/25$ was deemed a good compromise between simulation accuracy and efficiency. Thus, $a/25$ was adopted for all simulations with the X-intersection model. However, applying this high resolution to the entire DFN domain significantly increases computational demands~\citep[cf., ][]{matthai2024}. To address this issue for the variable-aperture model, we ensured that at least five finite-volume elements would span narrow fractures (i.e., $a \leq 2~\mathrm{mm}$). This resolution still sufficiently approximates flow profiles while $a/25$ was achieved for most fracture segments ( Fig.~\ref{fig:OH_mesh}b). Even with this relaxed strategy, $\sim$15 million elements were needed to discretise the fracture void space.

\begin{table}[h!]
	\centering
	\begin{tabular}{llcc}
		\hline
		\textbf{$u_0$} & \textbf{Mesh} & \textbf{$v_{max}$} & \textbf{$\Delta p$} \\
		\hline
		\multirow{3}{*}{$1~\mathrm{mm~s^{-1}}$} & $a/20$ & 0.01229 & 2178.7 \\
		& $a/25$ & 0.01225 & 2198.2 \\
		& $a/30$ & 0.01224 & 2203.2 \\
		\hline
		\multirow{3}{*}{$1~\mathrm{m~s^{-1}}$} & $a/20$ & 1.3609 & 14924 \\
		& $a/25$ & 1.3606 & 14919.1 \\
		&  $a/30$ & 1.3606 & 14917.1 \\
		\hline
	\end{tabular}
	\caption{\label{tab:mesh_conv}X-model: Variations in the maximum velocity and pressure drop across the fracture intersection with different mesh resolutions for fluid XG5000. Beyond a mesh resolution of $a/20$, the simulation results show only minor improvements.}
\end{table}

\begin{figure} 
	\centering
	\includegraphics[width=\columnwidth]{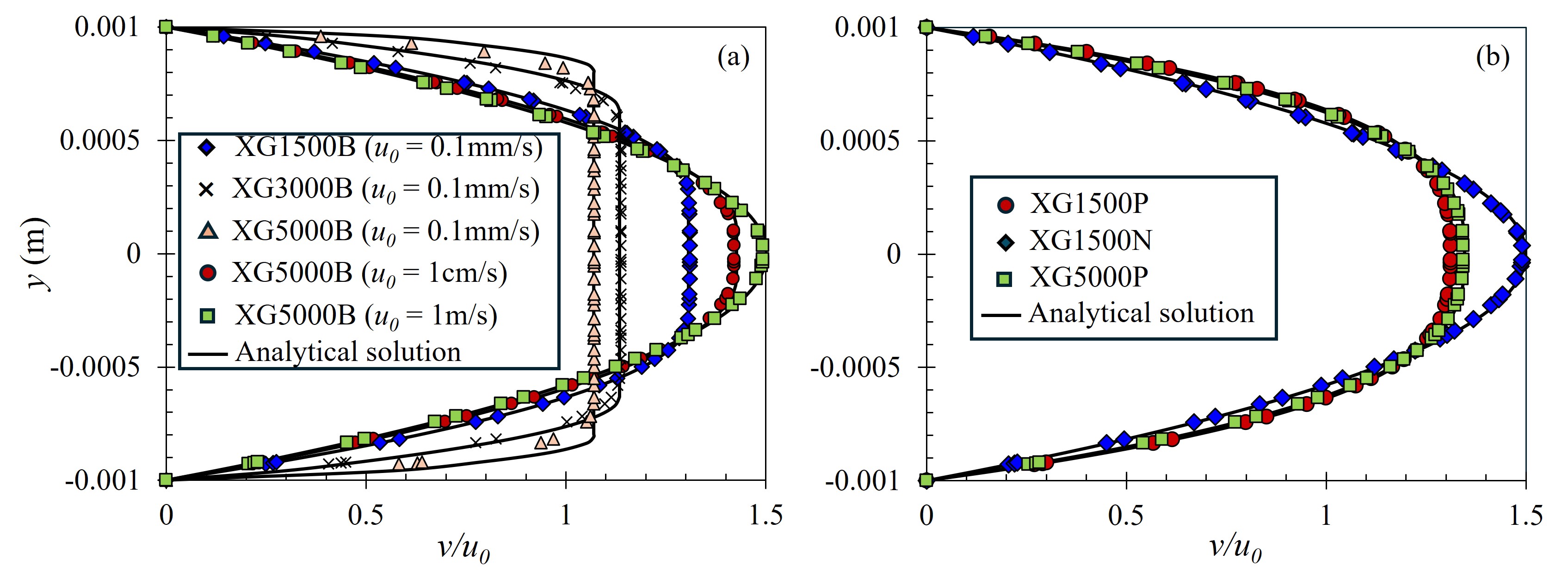}
	\caption{Comparison of fully developed velocity profiles for fracture-channel flow between numerical and analytical solutions for (a) Bingham fluids at various incoming velocities and (b) power-law fluids at $u_0 = 1~\mathrm{cm~s^{-1}}$. Simulation results perfectly match the analytical solutions.}
	\label{fig:Xmodel_valid}
\end{figure}

Model verification was performed separately for shear-thinning and yield stress properties, comparing simulated fully developed channel-flow profiles of single-rheology fluids with analytical solutions. We find that non-Newtonian velocity distributions significantly differ from their Newtonian counterparts (Fig.~\ref{fig:Xmodel_valid}). For Bingham fluids, the profiles exhibit a distinctive plateau at the centre of the fracture, known as the plug zone~\cite[][]{bird1983}, where variations in velocity
and thus shear rate are negligibly small (Fig.~\ref{fig:Xmodel_valid}a). This plateau expanded with increasing yield stress and decreasing fracture flow velocities. For shear-thinning power-law fluids, the parabolic profile flattened noticeably at the fracture core (Fig.~\ref{fig:Xmodel_valid}b). Both non-Newtonian features reduced the maximum velocity across the fracture aperture compared to the Newtonian fluid. 

Simulation results match the analytical solutions exactly for both Bingham~\cite[][]{bird1983} and power-law fluids~\cite[][]{tanner2000} across varying flow velocities and degrees of rheological effect (Fig.~\ref{fig:Xmodel_valid}). This confirms the reliability of our rheological modelling and finite-volume calculations.

\subsection{\label{sec:ssec3.2}Non-Newtonian effects on fracture flow across a single intersection}

To analyse the impact of a non-Newtonian rheology, influences of yield stress and shear-thinning behaviour are examined separately, identifying their individual contributions to fracture flow pattern and distribution. For the former, simulations are conducted using fluid XG5000B with varying $\tau_0$. The latter is analysed using fluid XG5000P with different $n$ values. The incoming velocity was varied from $u_0 = 0.1$~$\mathrm{mm~s^{-1}}$ to 10~$\mathrm{m~s^{-1}}$, covering a wide spectrum of fracture flow regimes.

\begin{figure} 
	\centering
	\includegraphics[width=\columnwidth]{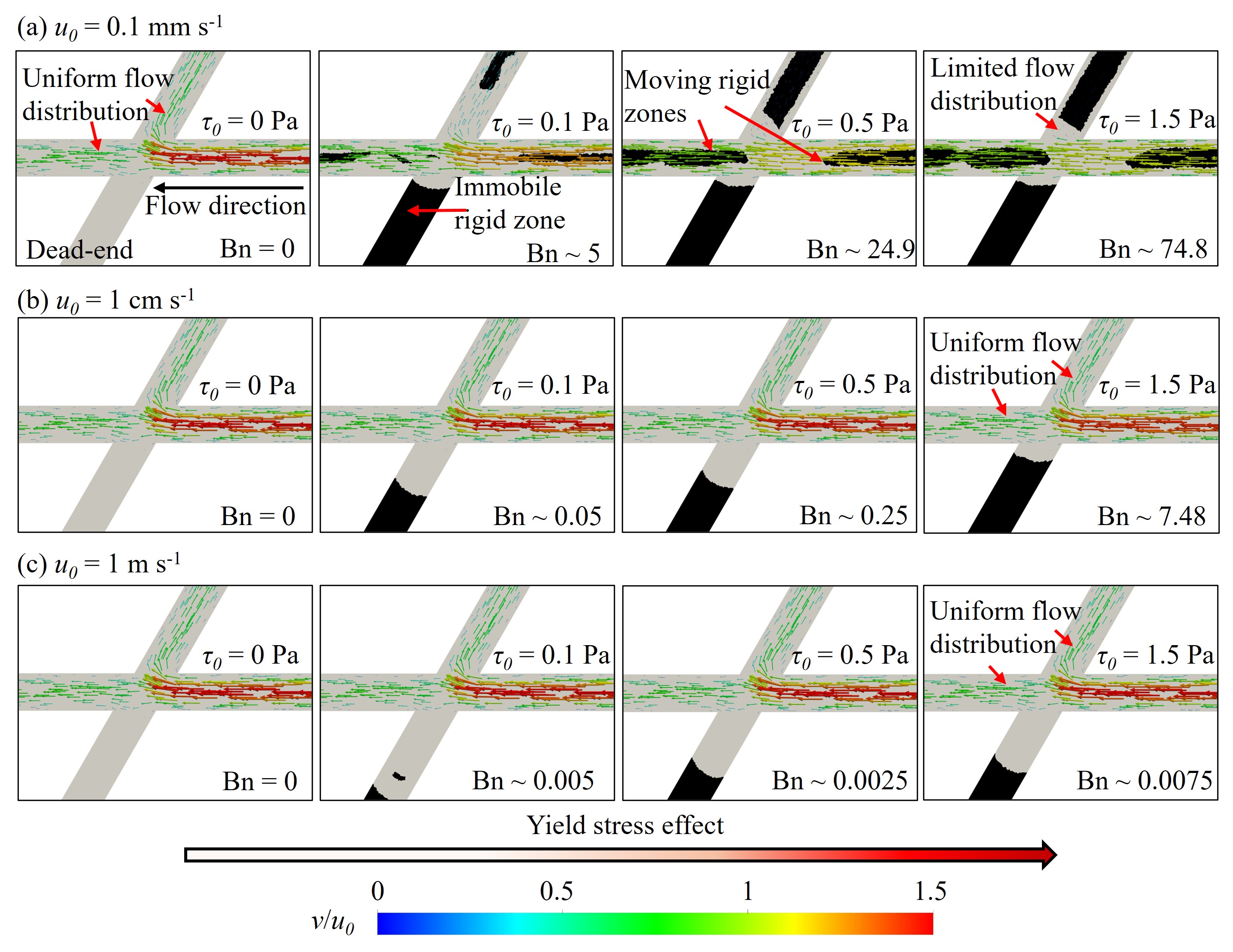}
	\caption{X-model: Velocity vector field and formation of rigid zones for fluid XG5000B with various values of yield stress ($\tau_0$) at (a) $u_0 = 10^{-4}~\mathrm{m~s^{-1}}$ ($\mathrm{Re} \sim 5 \times 10^{-4} $), (b) $u_0 = 10^{-1}~\mathrm{m~s^{-1}}$ ($\mathrm{Re} \sim 0.5$) and (c) $u_0 = 1~\mathrm{m~s^{-1}}$ ($\mathrm{Re} \sim 5$). As the yield stress increases, rigid zones (black) expand within fractures.}
	\label{fig:Xmodel_tau}
\end{figure}
\begin{figure} 
	\centering
	\includegraphics[width=\columnwidth]{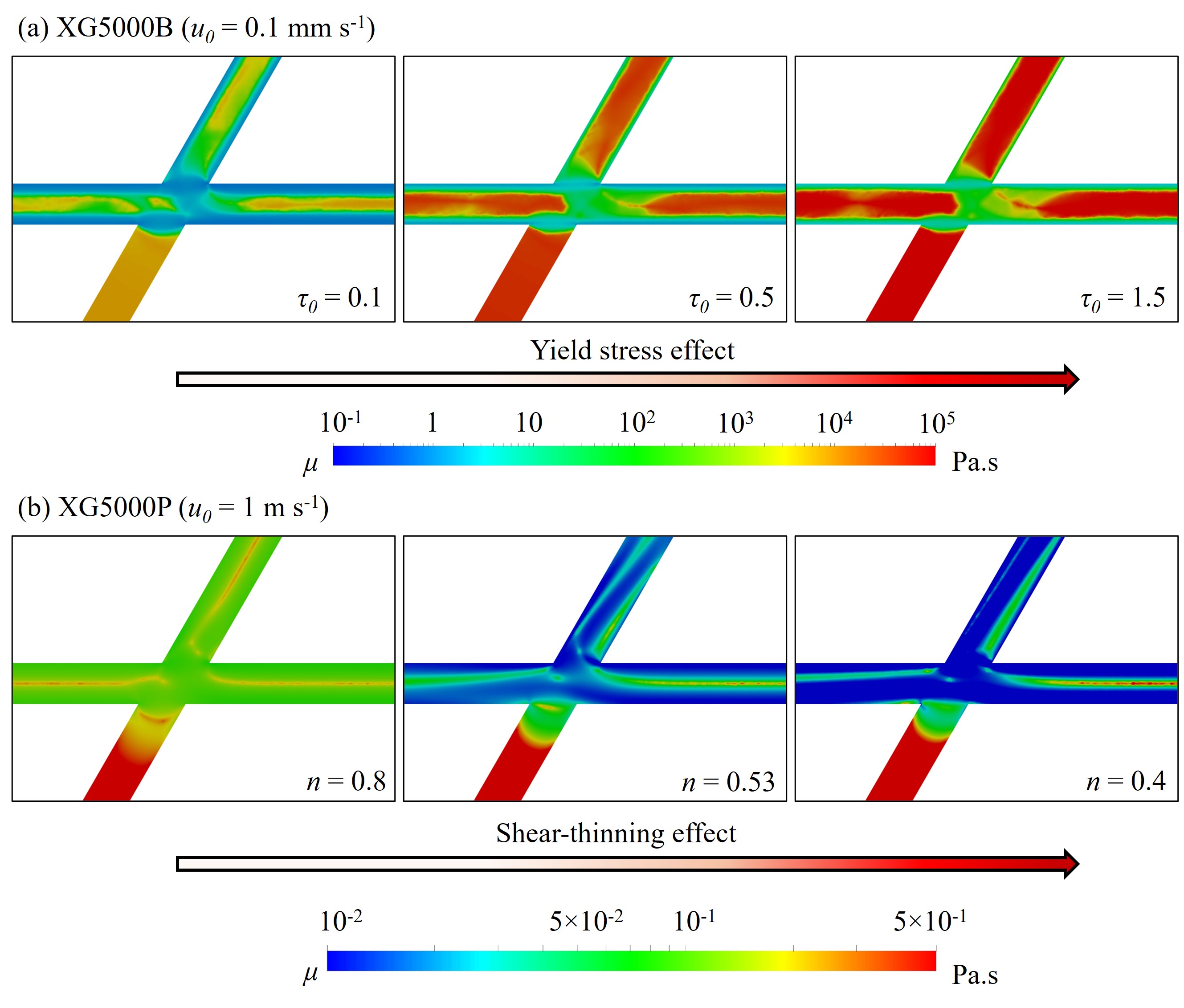}
	\caption{X-model: Snapshots of the viscosity field across the fracture intersection for (a) fluid XG5000B at $u_0 = 10^{-4}~\mathrm{m~s^{-1}}$ ($\mathrm{Re} \sim 5 \times 10^{-4} $) and (b) fluid XG5000P at $u_0 = 1~\mathrm{m~s^{-1}}$ ($\mathrm{Re} \sim 5$). Yield stress increases the apparent viscosity, whereas shear-thinning reduces it.}
	\label{fig:Xmodel_vis}
\end{figure}

For yield-stress-only XG5000B fluids, the Reynolds number spans from $\mathrm{Re} = 5 \times 10^{-4}$ to 50, and the Bingham number ranges from $\mathrm{Bn} = 5 \times 10^{-4}$ to 74.8, corresponding to the aforementioned range of inlet velocity and yield stresses of $\tau_0 = 0.1-1.5~\mathrm{Pa}$. Figure~\ref{fig:Xmodel_tau} illustrates the resulting variations in fracture flow patterns under these conditions. At a low velocity, i.e., $u_0 = 0.1~\mathrm{mm~s^{-1}}$ ($\mathrm{Re} \sim 5 \times 10^{-4}$), the flow-induced shearing is insufficient to fully yield the non-Newtonian fluid, resulting in the development of unyielded domains throughout the fracture intersection. These regions exhibit an extremely high effective viscosity, exceeding $10^5~\mathrm{Pa~s}$ (Fig.~\ref{fig:Xmodel_vis}a), thus behaving like a solid. Such domains can be immovably stuck at the dead-end branch, where the fluid remains stagnant with zero shear rate. Alternatively, they can form in the middle of  fracture channels where velocity gradients vanish. Unlike stagnant regions, these rigid zones can move with the flow, maintaining their solid-like structure while being transported by the surrounding fluid. This behaviour explains the plug-zone plateau observed in the fracture velocity profiles (Fig.~\ref{fig:Xmodel_valid}a).

\begin{figure} 
	\centering
	\includegraphics[width=\columnwidth]{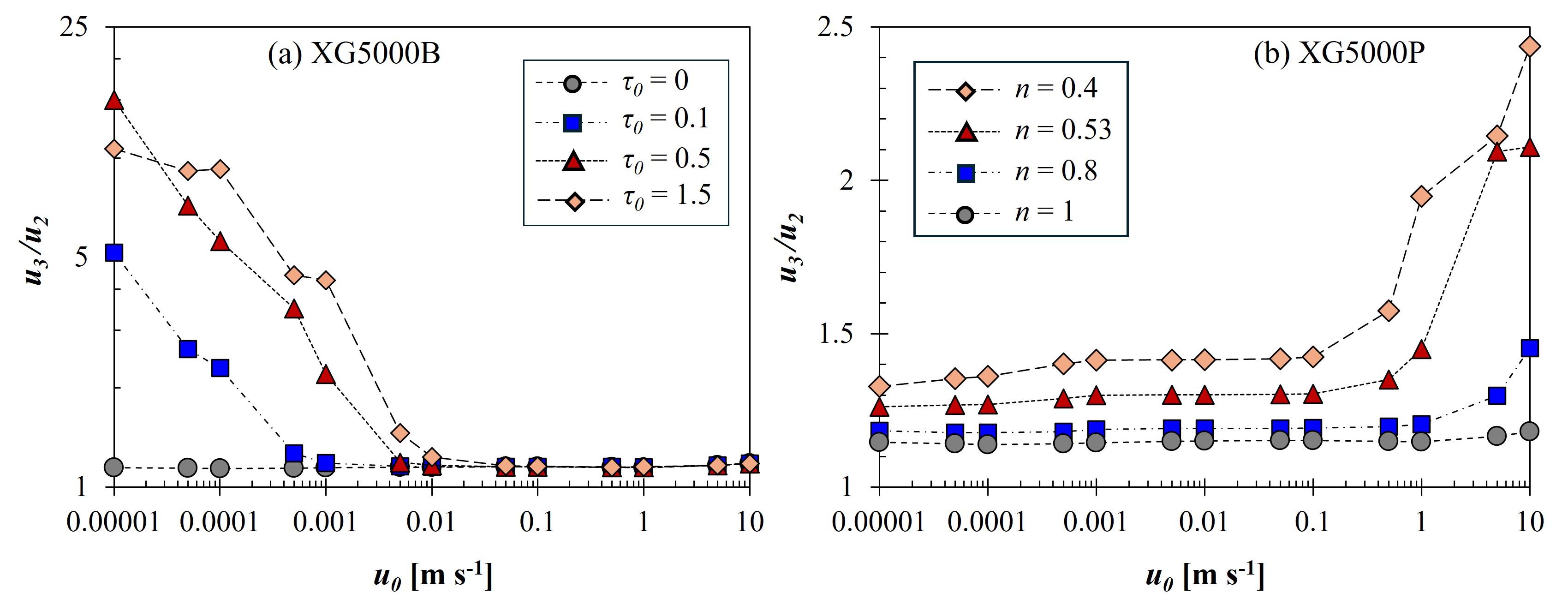}
	\caption{X-model: Flow partitioning at intersection (Fig.~\ref{fig:Xmodel_geometry}) for (a) fluid XG5000P with varying power-law indices ($n$) and (b) fluid XG5000B with different yield stress values ($\tau_0$). For XG5000P, as $n$ decreases, shear-thinning becomes more pronounced. For XG5000B, as $\tau_0$ increases, the yield stress effect gets more significant.}
	\label{fig:Xmodel_partitioning}
\end{figure}

As the yield stress increases (i.e., as $\mathrm{Bn}$ increases), a higher shear stress is required to initiate flow, leading to the expansion of unyielded rigid zones (Fig.~\ref{fig:Xmodel_tau}). At $u_0 = 0.1~\mathrm{mm~s^{-1}}$ and $\tau_0 \geq 0.5~\mathrm{Pa}$ ($\mathrm{Re} \sim 5 \times 10^{-4}$, $\mathrm{Bn} \geq 24.9$), the rigid zone in the side branch expands to occupy almost the entire fracture, seriously obstructing flow distribution into that branch. It is noteworthy that at $u_0 < 1~\mathrm{cm~s^{-1}}$ ($\mathrm{Re} < 5 \times 10^{-2}$), yielding occurs predominantly in the straight pressure gradient aligned fractures, with minimal ability to change direction, as indicated by the significantly greater flow partitioning into these segments (Fig.~\ref{fig:Xmodel_partitioning}a). At even higher velocities, i.e., $u_0 \geq 1~\mathrm{cm~s^{-1}}$ ($\mathrm{Re} \geq 5 \times 10^{-2}$), moving rigid zones disappear (Fig.~\ref{fig:Xmodel_tau}b, c). Given that the remaining dead zones do not contribute to the main flow, the flow patterns in active fractures closely resemble those of a Newtonian fluid at the same velocity, uniformly distributing fluid into branches (Fig.~\ref{fig:Xmodel_partitioning}a). Under these high-shear conditions, the effect of yield stress becomes negligible, thus leading to a Newtonian parabolic velocity profile at $u_0 = 1~\mathrm{m~s^{-1}}$ (Fig.~\ref{fig:Xmodel_valid}a).

\begin{figure} 
	\centering
	\includegraphics[width=\columnwidth]{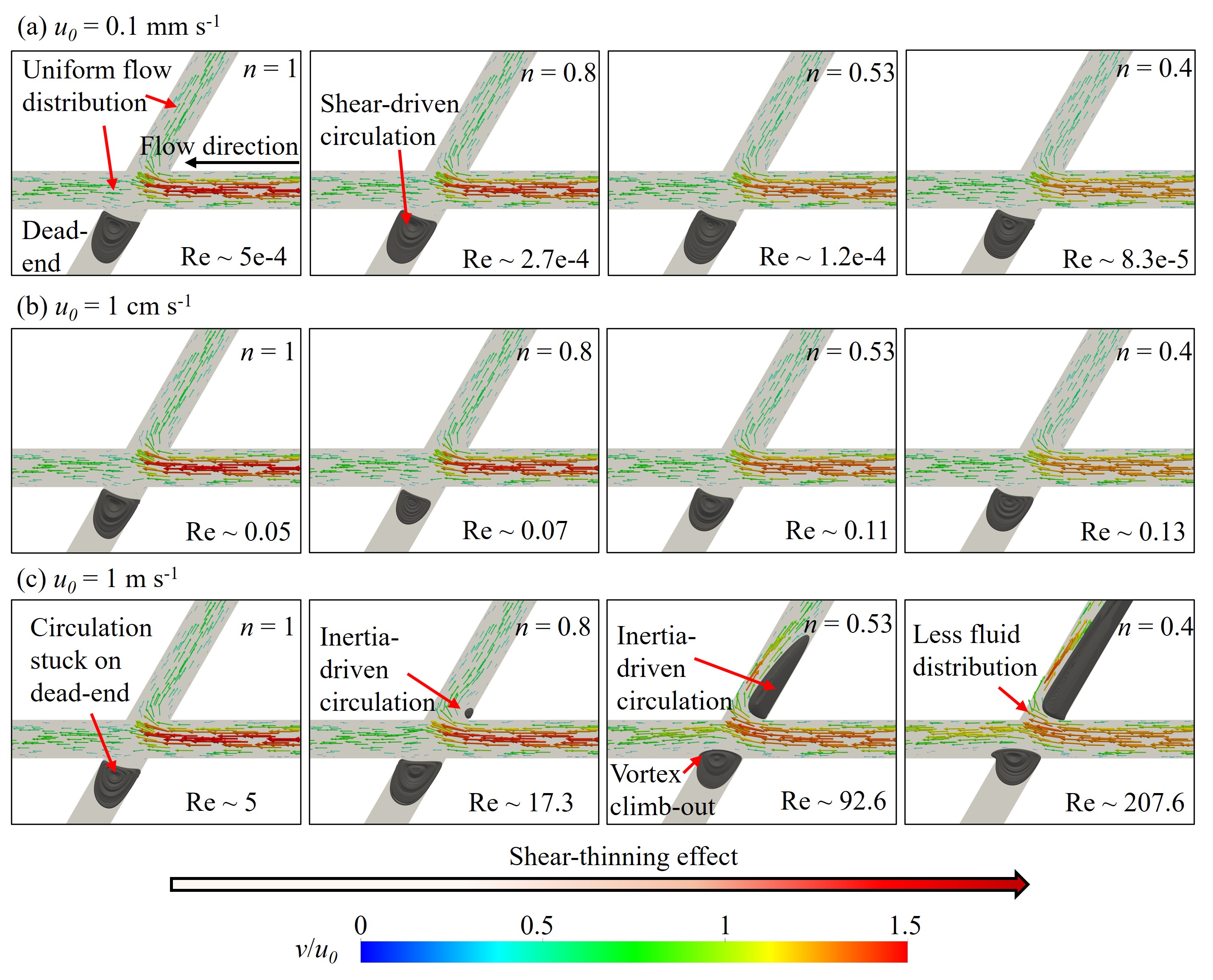}
	\caption{X-model: Velocity vector field and formation of recirculation zones for fluid XG5000P with various values of the shear-dependent power-law index ($n$) at (a) $u_0 = 10^{-4}~\mathrm{m~s^{-1}}$, (b) $u_0 = 10^{-1}~\mathrm{m~s^{-1}}$ and (c) $u_0 = 1~\mathrm{m~s^{-1}}$. Reynolds numbers corresponding to each combination of $n$ and $u_0$ are shown. As shear-thinning increases, circulation wakes expand.}
	\label{fig:Xmodel_n}
\end{figure}

Figure~\ref{fig:Xmodel_n} depicts variations in the flow structures of shear-thinning-only fluids XG5000P ($\mathrm{Bn} = 0$) as a function of the shear-thinning factor $n$. Over the investigated velocity range, the Reynolds number varies from $\mathrm{Re} = 8.3 \times 10^{-5}$ to 207.6 (for $n = 0.4$), indicating that shear-thinning controls the fluid inertia effect. At $u_0 \leq 1~\mathrm{cm~s^{-1}}$ ($\mathrm{Re} \leq 0.13$), Newtonian and non-Newtonian fluid flows exhibit similar patterns (Fig.~\ref{fig:Xmodel_n}a, b). A shear-driven circulation forms in the dead-end branch (cavity-like) but does not affect the overall flow in active fractures; this geometry-induced circulation remains approximately size-invariant as velocity (and Re) increases. At a high velocity, i.e., $u_0 = 1~\mathrm{m~s^{-1}}$, another wake is formed near the fracture intersection for $n = 0.53$ ($\mathrm{Re} = 92.6$) and 0.4 ($\mathrm{Re} = 207.6$) (Fig.~\ref{fig:Xmodel_n}c), resulting from the Bernoulli underpressure effect that reverses the side-branch flow into fast-flowing straight fractures~\citep[][]{matthai2024}. As $n$ decreases, the shear-thinning effect intensifies, and this recirculation zone enlarges and occupies a larger volume at the side branch entrance. Additionally, the shear-driven circulation extends out of the dead-end branch, influencing fracture flow by bending its bulk trajectory. These changes occur because, at high velocities, the large shear rate reduces the effective viscosity of the shear-thinning fluid by up to two orders of magnitude, thereby significantly amplifying inertia effects, as evidenced by the pronounced increase in $\mathrm{Re}$ (Fig.~\ref{fig:Xmodel_vis}b). Secondary swirling flows confine fracture parallel flow to narrower regions (Fig.~\ref{fig:Xmodel_n}c), enhancing flow channeling. This had been observed previously by \cite[][]{lavrov2013redirection} and \cite[][]{zhang2023displacement}. 

Furthermore, shear-thinning significantly alters the flow distribution, increasingly favouring the straight pathway as $n$ decreases (Fig.~\ref{fig:Xmodel_partitioning}b). At $u_0 > 0.1~\mathrm{m~s^{-1}}$, a remarkable jump in flow into the straight effluent fracture occurs, likely due to the enlargement of the recirculation zone in the side branch (Fig.~\ref{fig:Xmodel_n}c). This further distorts the flow partitioning, reinforcing the characteristic channeling of shear-thinning fluids.

\begin{figure} 
	\centering
	\includegraphics[width=\columnwidth]{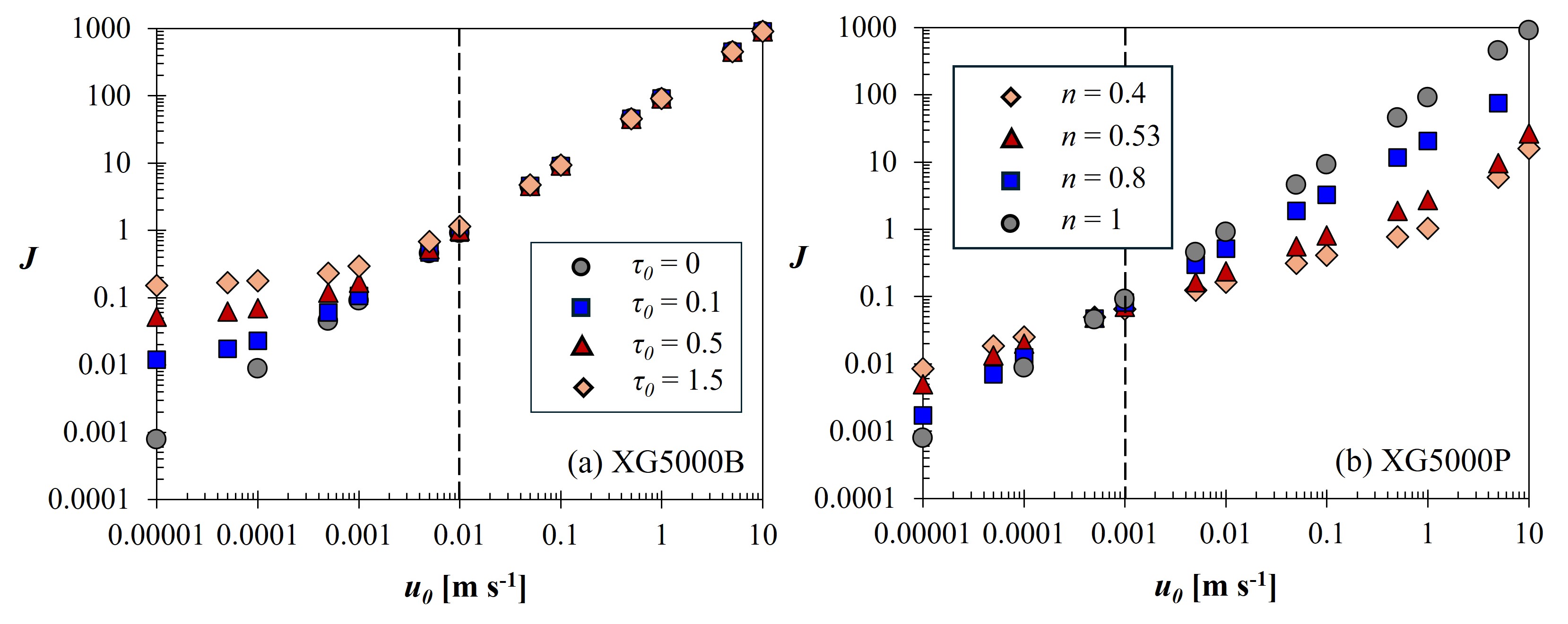}
	\caption{X-model: Hydraulic gradient, $J$ across the X-intersection (Fig.~\ref{fig:Xmodel_geometry}) for (a) XG5000P with varying power-law indices ($n$) and (b) XG5000B with different yield stress values ($\tau_0$).}
	\label{fig:Xmodel_J}
\end{figure}

For Bingham fluids, the hydraulic gradient, $J$, across the fracture intersection changes significantly with the flow velocity (Fig.~\ref{fig:Xmodel_J}a). At low inlet velocities ($u_0 < 1~\mathrm{cm~s^{-1}}$), the larger yield stress results in a higher $J$, implying greater pressure losses. The fluid requires additional energy to yield and overcome the resistance of rigid zones. At higher velocities ($u_0 \geq 1~\mathrm{cm~s^{-1}}$), the influence of yield stress diminishes as the fluid becomes fully yielded, resulting in $J$ values throughout the active fractures similar to those of a Newtonian fluid. No additional sensitivity to energy losses is seen. Shear-thinning increases the hydraulic gradient at $u_0 < 1~\mathrm{mm~s^{-1}}$ (Fig.~\ref{fig:Xmodel_J}b). As the fluid becomes thicker under these low-shear conditions, larger viscous losses occur. This trend reverses at $u_0 \geq 0.1~\mathrm{mm~s^{-1}}$, where the reduced effective viscosity leads to decreased energy losses.

In summary, complex non-Newtonian rheology fundamentally alters fracture flow, introducing rigid zones and inertia-induced circulations as well as shifting flow partitioning and pressure losses across the single X-shaped fracture intersection.


\subsection{Non-Newtonian effects on fracture flow in the natural network}

The flow of non-Newtonian xanthan biopolymer solutions (XG1500, XG3000, and XG5000) is simulated with the Hornelen network model for constant injection rates ranging from $10^{-5}$ to $10^{-2}~\mathrm{m^3 m^{-2} s^{-1}}$. For these conditions, we evaluate the combined effects of shear-thinning and yield stress rheology. This analysis encompasses shear-rate and viscosity distributions, local flow structures, network flow partitioning, and the overall pressure drop across the network, considering the geometrical complexities imposed by natural fracture patterns.

\subsubsection{Shear-rate and viscosity distributions}

\begin{figure} 
	\centering
	\includegraphics[width=\columnwidth]{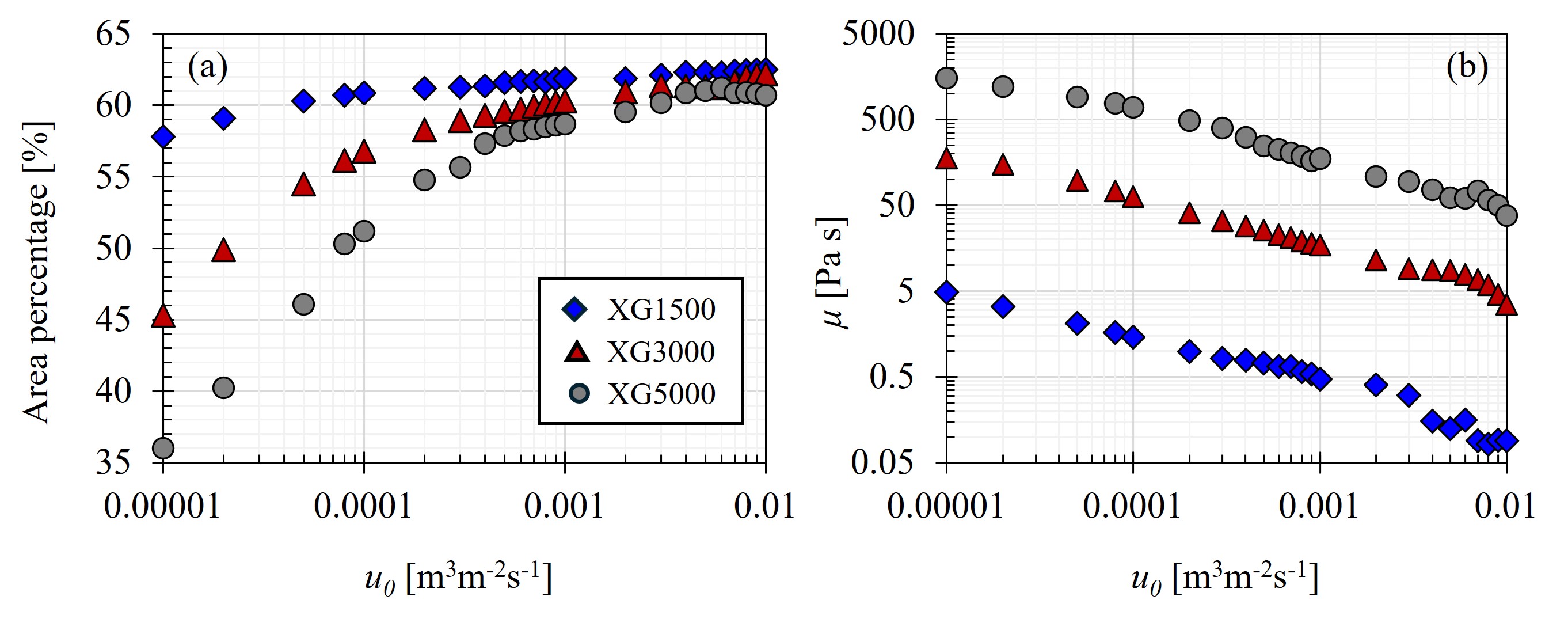}
	\caption{Hornelen network model: Area percentage of the yielded zone created by xanthan polymer solutions in the fracture network as a function of injection rate. Fluids XG3000 and XG5000 yield in only $\leq 60\%$ of the fracture area at $u_0 \leq 10^{-3}~\mathrm{m^3 m^{-2} s^{-1}}$.}
	\label{fig:OH_avg-mu&yielded}
\end{figure}

At low injection rates, the flowing region (yielded zone) occupies a limited portion of the fracture area, covering only $\sim 45\%$ for XG3000 and $\sim 36\%$ for XG5000 at $u_0 = 10^{-5}~\mathrm{m^3 m^{-2} s^{-1}}$ (Fig.~\ref{fig:OH_avg-mu&yielded}a). This occurs because the velocity gradient is too low to overcome the yield stress in much of the network, particularly in regions with lower permeability or reduced flow connectivity. This yielded zone expands progressively with increasing flow rate, reaching $\sim 60\%$ at $u_0 = 10^{-2}~\mathrm{m^3 m^{-2} s^{-1}}$ for all fluids.

\begin{figure} 
	\centering
	\includegraphics[width=\columnwidth]{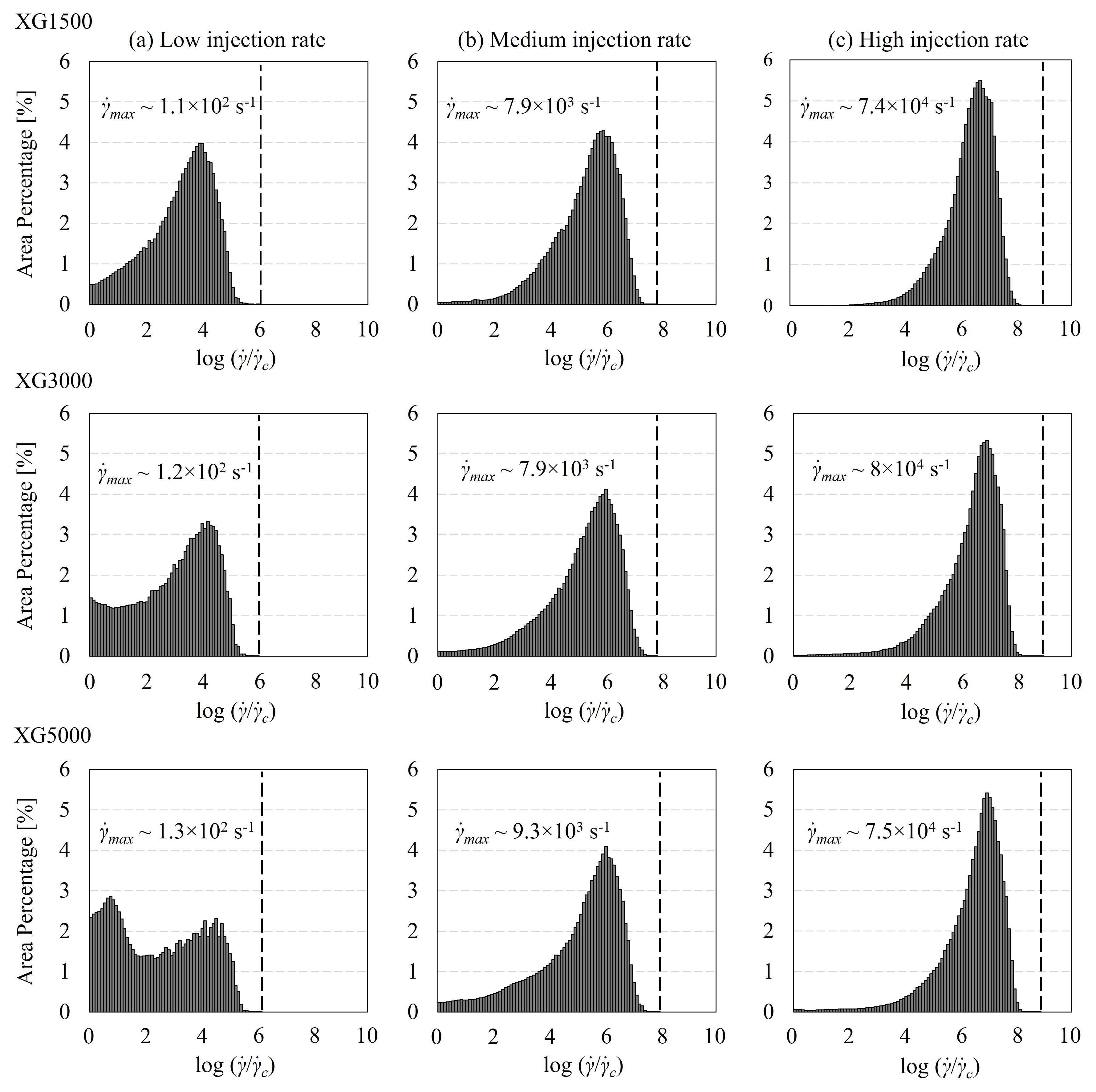}
	\caption{Hornelen network model: Logarithmic histograms of shear rate, normalised by the critical shear rate ($\dot{\gamma_c}$) for different fluids at (a) $u_0 = 10^{-5}$, (b) $10^{-3}$ , and (c) $10^{-2}~\mathrm{m^3 m^{-2} s^{-1}}$. Unyielded rigid zones are excluded. As $u_0$ increases, the shear-rate range shifts to higher values, spanning $\sim 9$ orders of magnitude above $\dot{\gamma_c}$.}
	\label{fig:OH_shear_hist}
\end{figure}

\begin{figure} 
	\centering
	\includegraphics[width=\columnwidth]{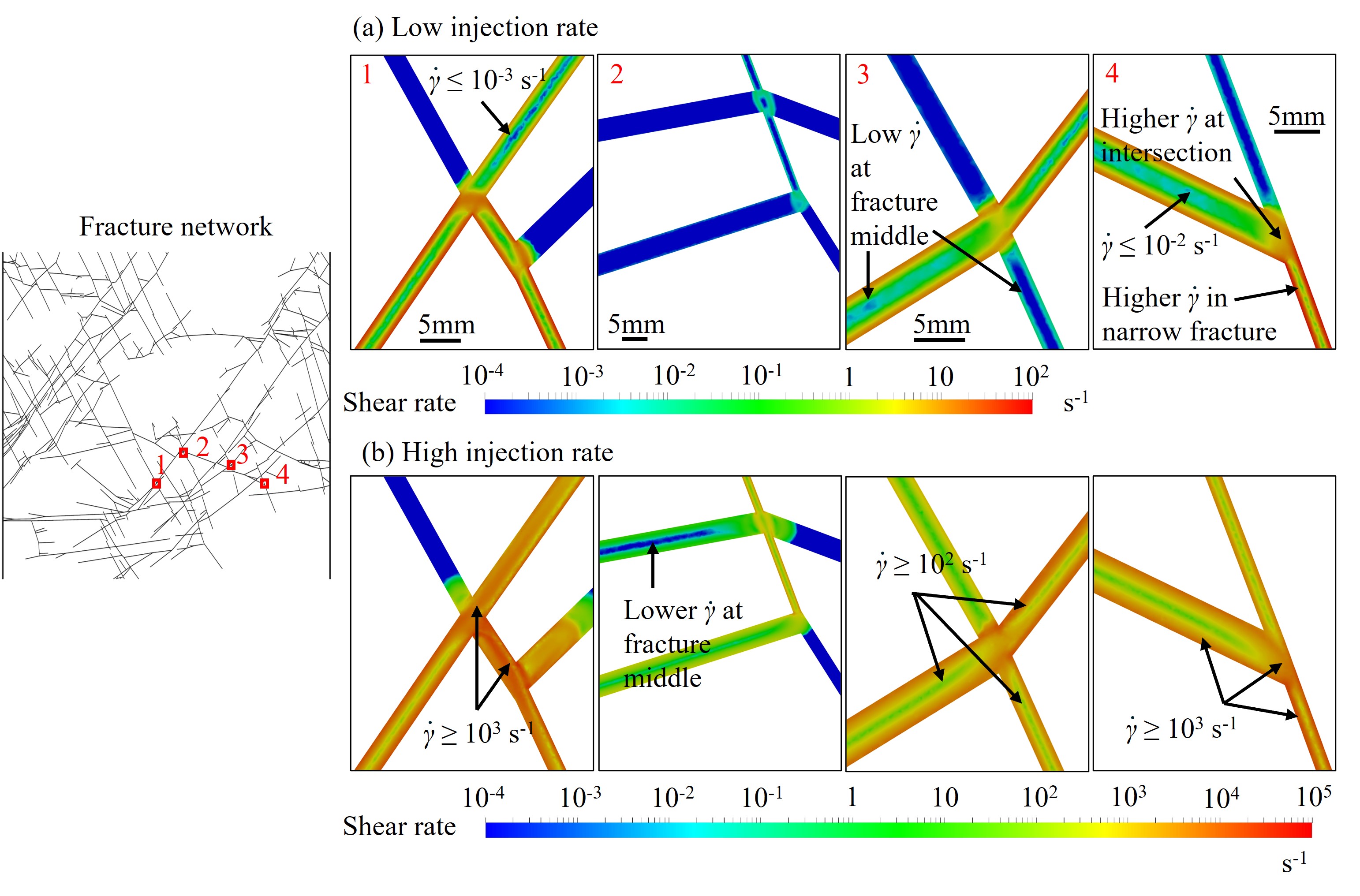}
	\caption{Hornelen network model: Shear rate distribution for fluid XG5000 at (a) $u_0 = 10^{-5}$, and (b) $10^{-2}~\mathrm{m^3 m^{-2} s^{-1}}$.}
	\label{fig:OH_shear_contour}
\end{figure}

\begin{figure} 
	\centering
	\includegraphics[width=\columnwidth]{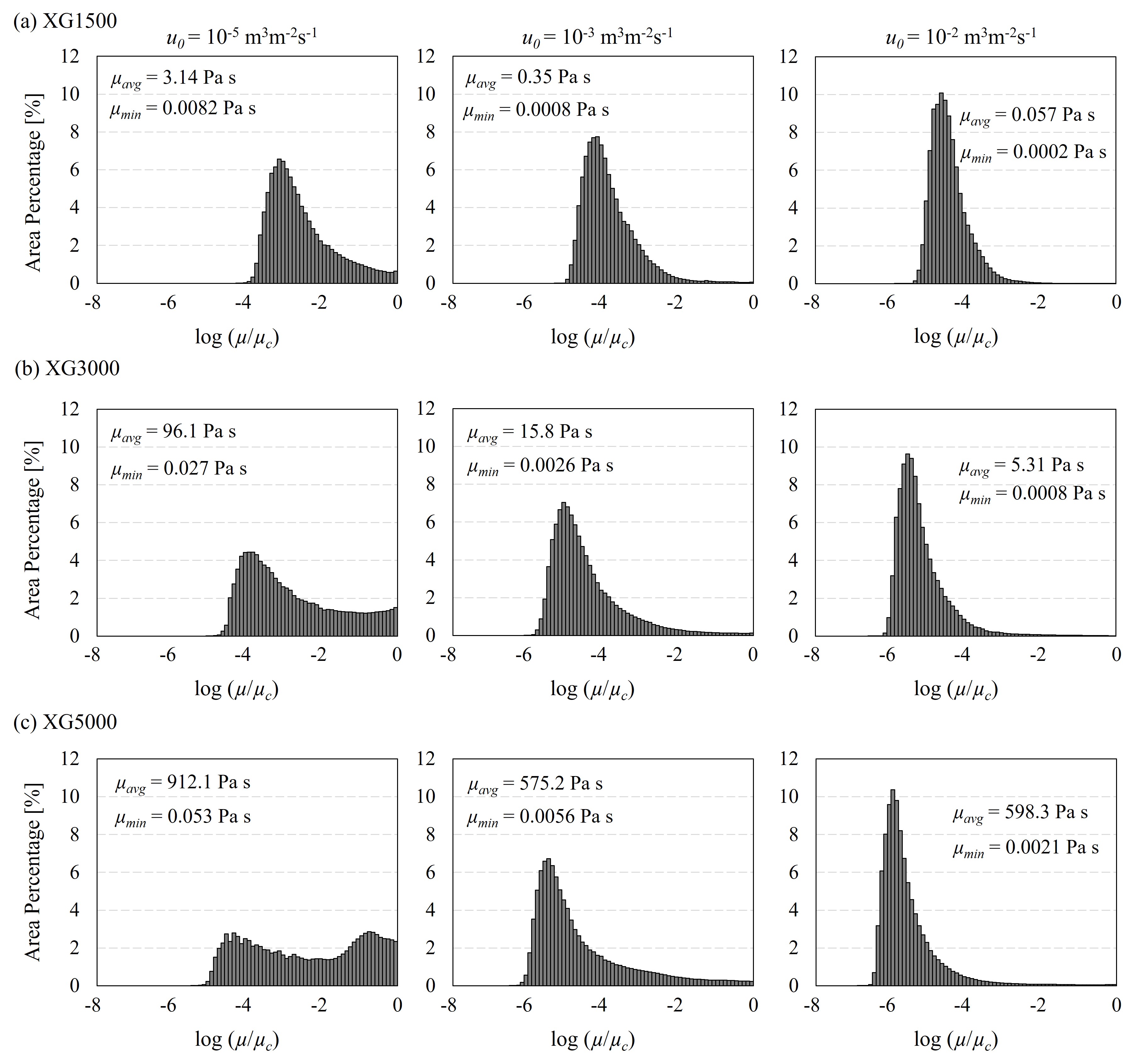}
	\caption{Hornelen network model: Logarithmic histograms of apparent viscosity normalised by the critical value ($\mu/\mu_c$), for different fluids at (a) $u_0 = 10^{-5}$, (b) $10^{-3}$, and (c) $10^{-2}~\mathrm{m^3 m^{-2} s^{-1}}$. $\mu_{avg}$ and $\mu_{min}$ denote the network-average and minimum over yielded fractures; unyielded rigid zones are omitted. The critical (pre$-$shear-thinning) viscosities are $\mu_{c} = 150$, $3 \times 10^3$, and $1.5 \times 10^4~\mathrm{Pa~s}$ for XG1500, XG3000, and XG5000, respectively.}
	\label{fig:OH_vishistogram}
\end{figure}

At $u_0 = 10^{-5}~\mathrm{m^3 m^{-2} s^{-1}}$, the shear rate is still small in much of the model even though the fluid has already yielded (Fig.~\ref{fig:OH_shear_hist}a). Shear rates below $0.01~\mathrm{s^{-1}}$ account for $\sim 17.1\%$, $\sim 25.6\%$, and $\sim 43.5\%$ of the fracture void space where fluids XG1500, XG3000, and XG5000 have yielded, respectively. This highlights the predominance of slow flow in wide fractures, where minimal geometrical confinement results in local velocity gradients (Fig.~\ref{fig:OH_shear_contour}a). The low-shear behaviour dominates fluid rheology, producing a large network-average viscosity (Fig.~\ref{fig:OH_avg-mu&yielded}b) and a broad distribution skewed toward large viscosity values (Fig.~\ref{fig:OH_vishistogram}a), with viscosities $\mu \geq 10~\mathrm{Pa \cdot s}$, over 10000 times that of water, representing up to $\sim 26.5\%$ of the yielded zone for fluid XG5000. At low shear rates (i.e., $\dot{\gamma} \leq 0.01~\mathrm{s^{-1}}$), the yield stress term, $\tau_0/\dot{\gamma}$, governs the apparent viscosity, as shown in the brackets of Eq.~(\ref{eqn_HB}). Although shear-thinning also increases fluid viscosity, its contribution is minimal compared to the yield stress effect, with fluid XG5000 exhibiting a viscosity exceeding $40~\mathrm{Pa~s}$ due to its yield stress property alone ($\tau_0 = 0.401~\mathrm{Pa}$). Consequently, the network-average viscosity of XG5000 is $\sim 314$ times higher than that of XG1500 for which the lower yield stress ($\tau_0 = 0.015~\mathrm{Pa}$) translates into a reduced resistance and a smaller proportion of high-viscosity areas (Fig.\ref{fig:OH_vishistogram}a).

Elevated injection rates display a broadening shear-rate distribution, extending $\sim 9$ orders of magnitudes above $\dot{\gamma_c}$, and a pronounced shift toward higher shear rates within the fracture network (Fig.~\ref{fig:OH_shear_hist}b, c). For fluid XG5000, shear rates exceeding $1~\mathrm{s^{-1}}$ becomes increasingly common, comprising $\sim 62.2\%$ of yielded fractures at $u_0 = 10^{-3}$ and $\sim 89.3\%$ at $10^{-2}~\mathrm{m^3 m^{-2} s^{-1}}$. This transition reflects larger local velocity gradients, generated by higher flow velocities. As a consequence, high-shear zones extend beyond narrow apertures and intersections into many parts of the network (Fig.~\ref{fig:OH_shear_contour}b). At these high shear rates, the shear-thinning term, $K {\dot{\gamma}}^{n-1}$, takes precedence in controlling viscosity (Eq.~(\ref{eqn_HB})), dropping it by more than two orders of magnitude from the critical viscosity, $\mu_{\mathrm{c}}$, across most of the network (Fig.~\ref{fig:OH_vishistogram}b, c). This critical viscosity corresponds to the yield point, where the fluid transitions from unyielded to yielded behaviour, and represents the maximum viscosity the fluid exhibits before shear-thinning effects dominate. Areas with a viscosity below $0.01~\mathrm{Pa~s}$ occupy a significant network portion, reaching $\sim 10.1\%$ of the yielded zone for XG5000 and $\sim 48\%$ for XG1500. The much larger proportion observed for fluid XG1500 is attributed to its lower $K$ and shear-thinning power-law index $n$, causing a steeper decline in the effective viscosity at high shear rates as compared with XG5000.

The large shear-rate variation within the fracture network leads to significant viscosity heterogeneity, spanning many orders of magnitudes, with the peak shifting from high values at low shear rates to much lower ones under high-shear stress conditions.

\subsubsection{Local flow pattern and network flow distribution}

Due to the geometric heterogeneity of the Hornelen fracture network, fluid flow exhibits diverse dynamic behaviours, transitioning from laminar creeping to inertia-dominated regimes~\citep[][]{matthai2024}. At any influx, the fracture flow velocity of fluid XG5000 can reach $v_{max} \sim 2000 u_0$, while the network-average velocity is only $v_{avg} \sim 110 u_0$. This indicates a substantial velocity variation across this network.

\begin{figure} 
	\centering
	\includegraphics[width=\columnwidth]{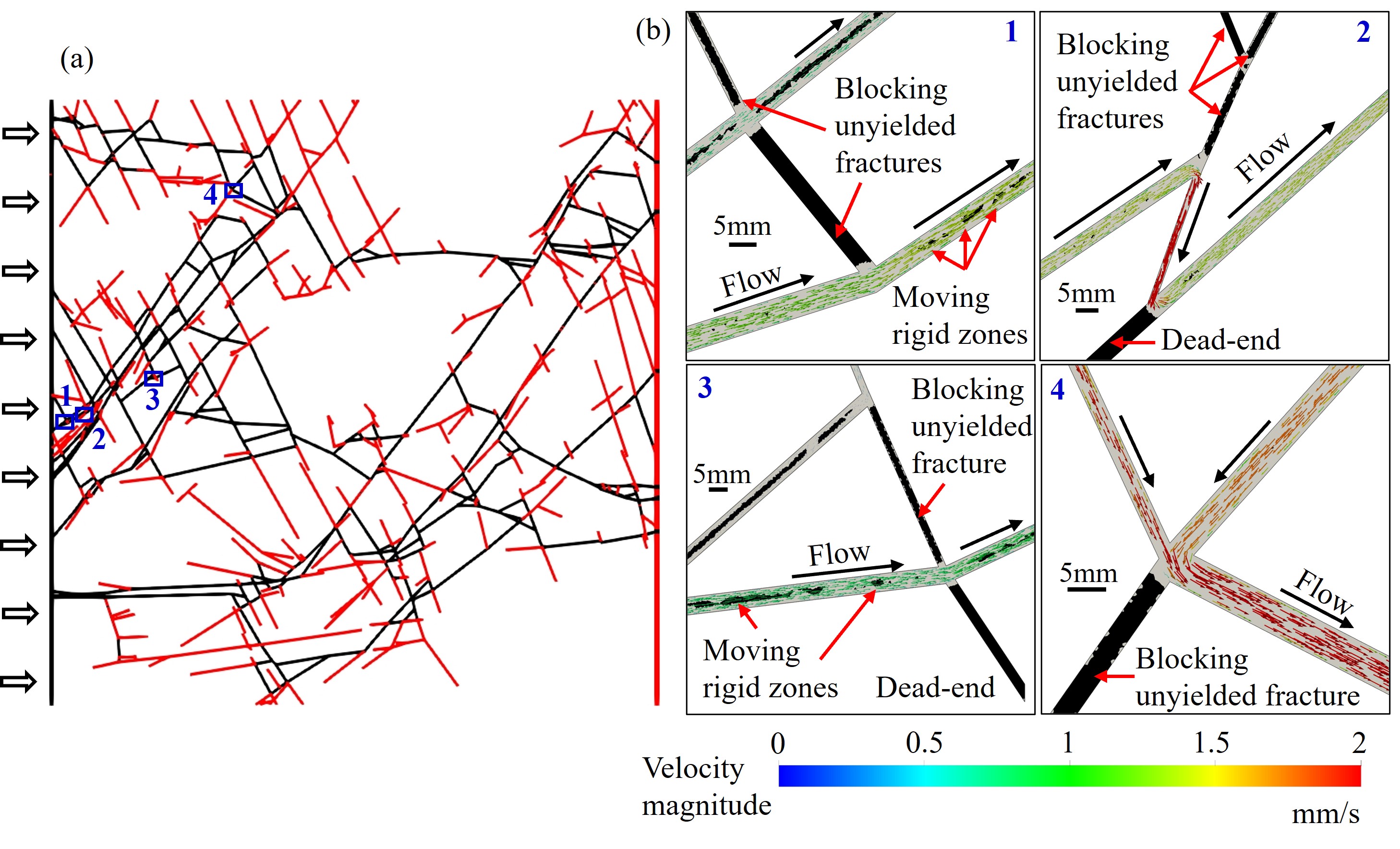}
	\caption{Hornelen network model: (a) Formation of unyielded fractures, highlighted in red, for fluid XG5000 at a low injection rate ($u_0 = 10^{-5}~\mathrm{m^3 m^{-2} s^{-1}}$). (b) A closer view of local flow patterns within typical fractures, showing the coexistence of immobile and moving rigid zones.}
	\label{fig:OH_rigid-defs}
\end{figure}

\begin{figure} 
	\centering
	\includegraphics[width=\columnwidth]{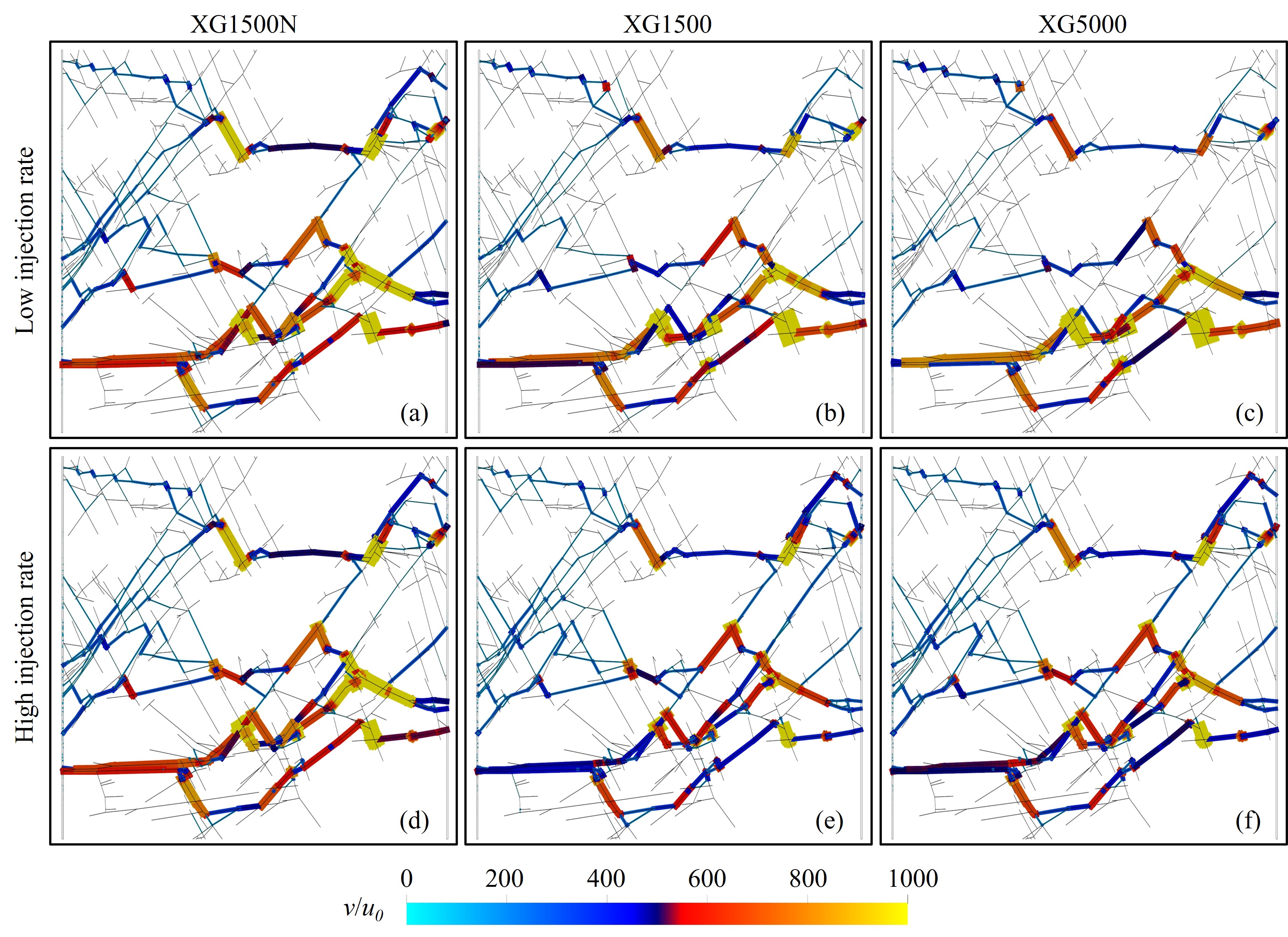}
	\caption{Hornelen network model: Network flow partitioning at a low influx ($u_0 = 10^{-5}~\mathrm{m^3 m^{-2} s^{-1}}$) and a high influx ($u_0 = 10^{-2}~\mathrm{m^3 m^{-2} s^{-1}}$) for different fluids.}
	\label{fig:OH_partitioning}
\end{figure}

A low influx of $u_0 = 10^{-5}~\mathrm{m^3 m^{-2} s^{-1}}$ results in millimetre-per-second fracture flow velocities and a dense distribution of unyielded zones occupying up to $\sim 65\%$ of the fracture cross-sectional area (Fig.~\ref{fig:OH_rigid-defs}). These dead-zones define the limited extent of the flowing region (Fig.~\ref{fig:OH_avg-mu&yielded}a). In addition to dead-ends, immobile rigid zones form within interconnected branches, completely blocking flow into these areas (Fig.~\ref{fig:OH_rigid-defs}b). Moving but rigid zones are prevalent, further impeding overall flow. This formation of rigid zones is governed by the yield stress rheology (Section~\ref{sec:ssec3.2}), significantly altering network flow at low injection rates (Fig.~\ref{fig:OH_partitioning}a-c). At the same influx, the Newtonian fluid XG1500N is distributed widely across the network, driven by Stokes flow, and the flow maintains a high network connectivity (Fig.~\ref{fig:OH_partitioning}a). By contrast, for fluid XG1500, with a yield stress ($\tau_0 = 0.015~\mathrm{Pa}$), flow partitioning is restricted to dominant pathways and many branches are deactivated (Fig.~\ref{fig:OH_partitioning}b). Within the flowing fractures, the flow velocity is reduced as compared to XG1500N due to the locally increased viscosity and moving rigid zones, adding extra resistance to the partially obstructed network flow. Fluid XG5000 ($\tau_0 = 1.5~\mathrm{Pa}$) features the most confined flow. Unyielded fractures limit the network flow to only four pathways, leaving much of the network inactive (Fig.~\ref{fig:OH_partitioning}c). This demonstrates that increasing yield stress enhances flow localisation, concentrating flow where shear stress is high.


\begin{figure} 
	\centering
	\includegraphics[width=\columnwidth]{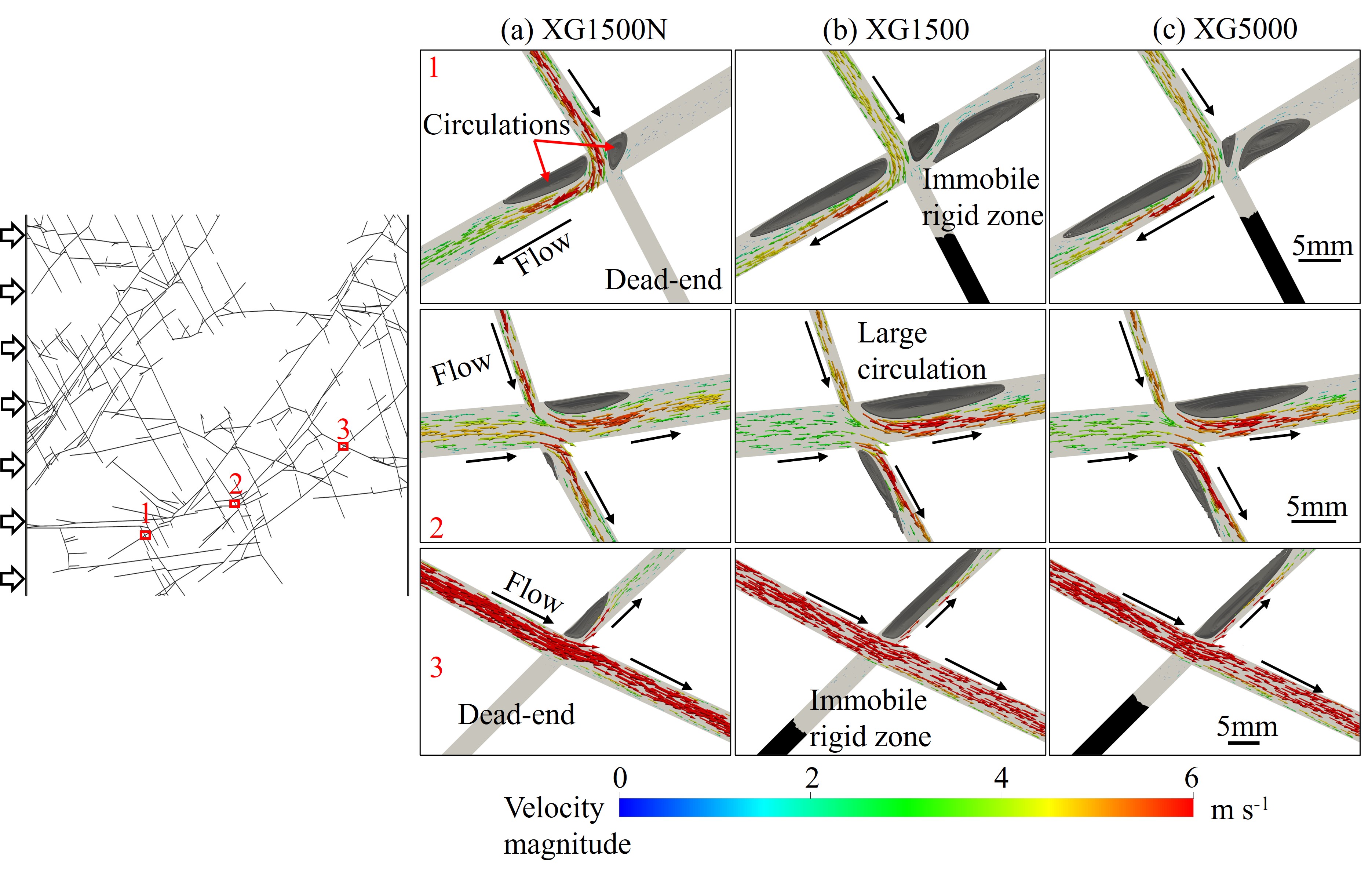}
	\caption{Hornelen network model: Flow patterns across fracture intersections at a high network influx ($u_0 = 10^{-2}~\mathrm{m^3 m^{-2} s^{-1}}$) for fluids XG1500N, XG1500, and XG5000. Non-Newtonian fluids (no 'N' in name) generate larger circulation bubbles compared to Newtonian fluid (XG1500).}
	\label{fig:OH_vortices_u1e-2}
\end{figure}

\begin{figure} 
	\centering
	\includegraphics[width=\columnwidth]{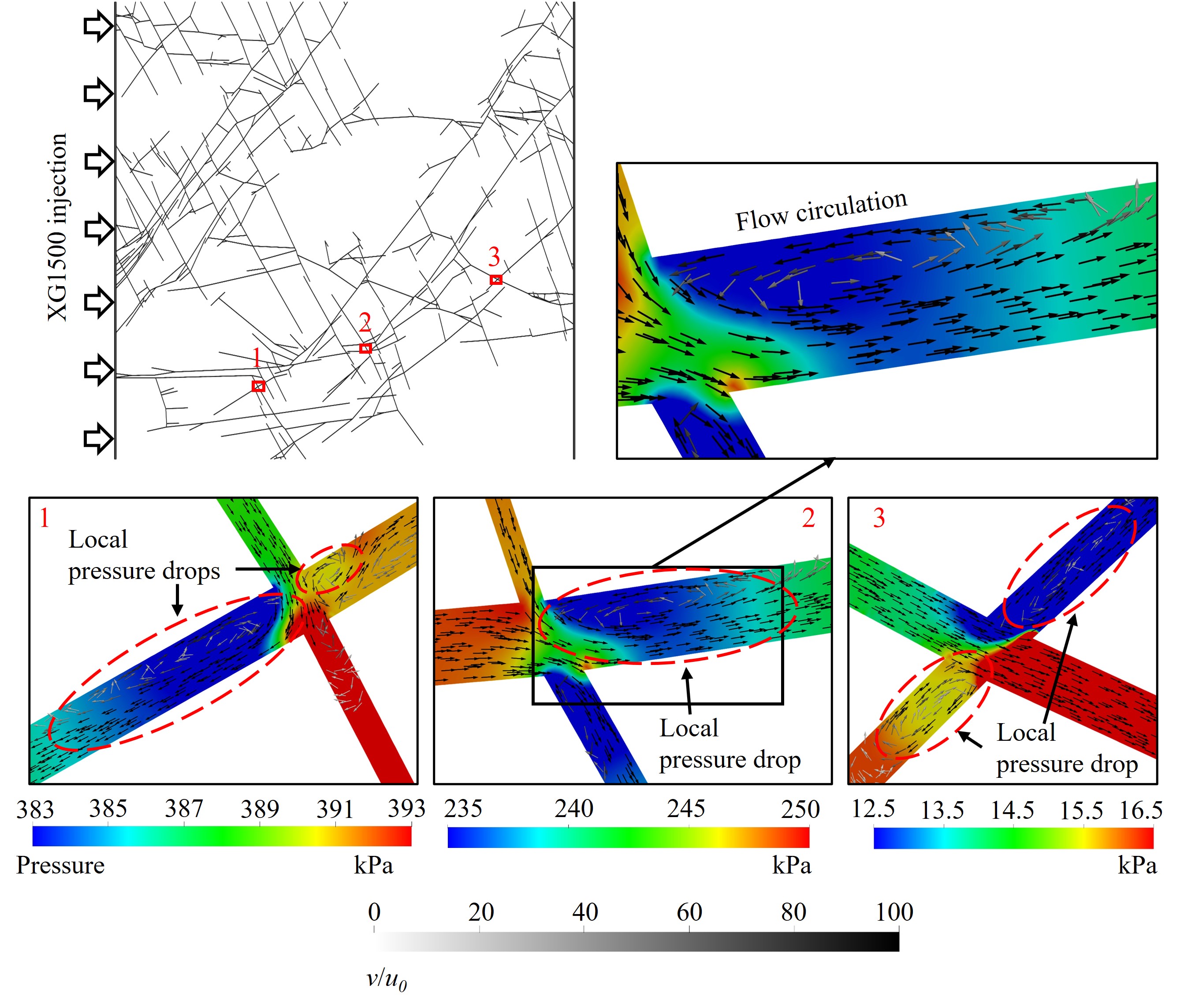}
	\caption{Hornelen network model: Pressure distribution across fracture intersections at a high network influx ($u_0 = 10^{-2}~\mathrm{m^3 m^{-2} s^{-1}}$) for fluid XG1500. Circulation formation results in local pressure losses near intersections.}
	\label{fig:OH_p_u1e-2}
\end{figure}

\begin{figure} 
	\centering
	\includegraphics[width=0.7\columnwidth]{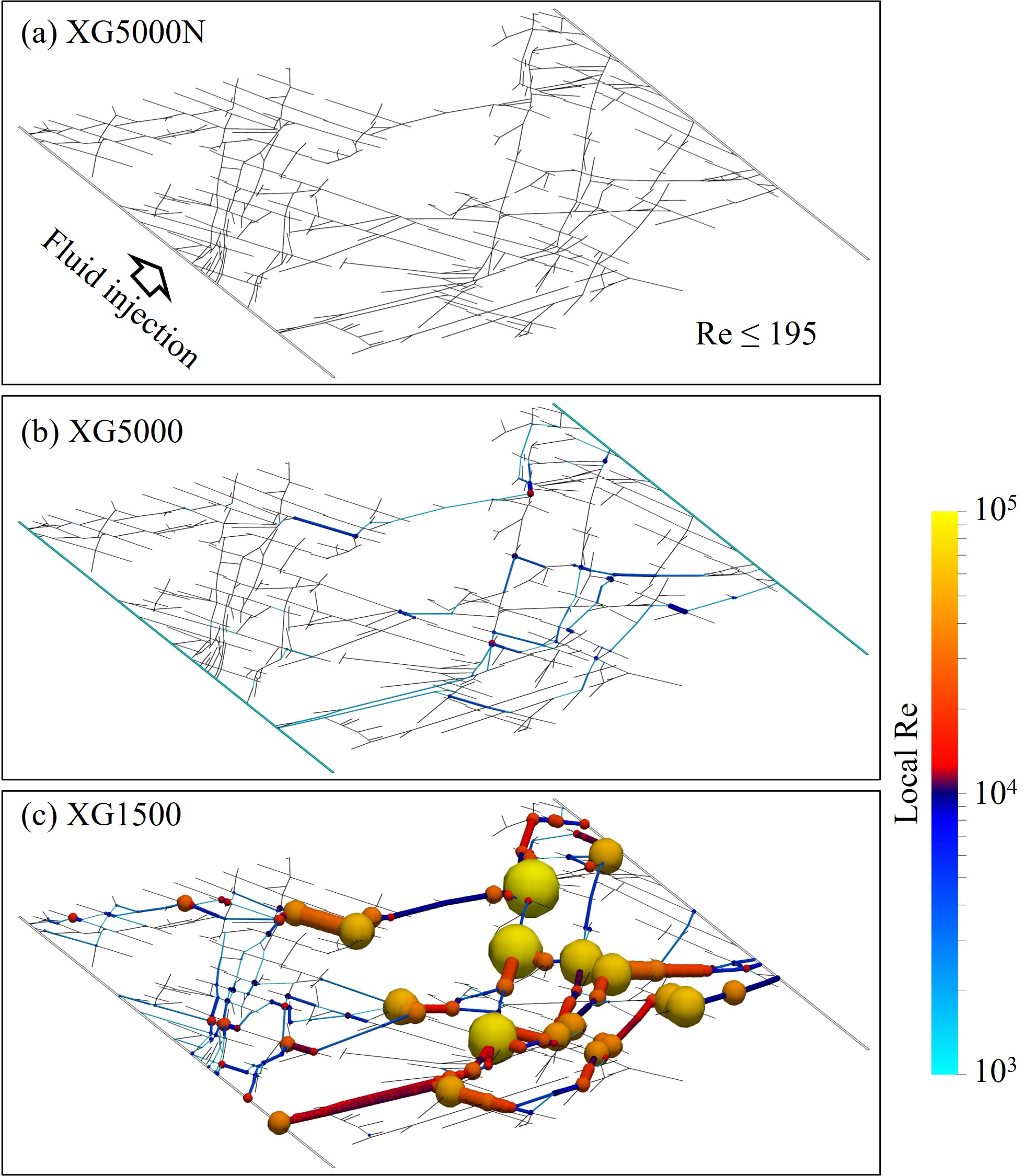}
	\caption{Hornelen network model: Local Reynolds number ($\mathrm{Re} = \frac{\rho v^{2-n} d_f^n}{\mu_{app}}$) at $u_0 = 10^{-2}~\mathrm{m^3 m^{-2} s^{-1}}$ for fluids (a) XG5000N, (b) XG5000, and (c) XG1500. Local $\mathrm{Re}$ is visualised with spherical glyphs; glyph size and colour scale with magnitude. Non-Newtonian cases show hotspots with $\mathrm{Re} \geq 10^4$, indicating a turbulent flow regime.}
	\label{fig:OH_Re_u1e-2}
\end{figure}

For a high fluid influx ($u_0 = 10^{-2}~\mathrm{m^3 m^{-2} s^{-1}}$), fracture flow reaches velocities in the metre-per-second range (Fig.~\ref{fig:OH_vortices_u1e-2}). The rigid zones disappear because the shear rate is sufficient to yield the fluid throughout the network, except in dead-end branches. In fracture intersections, the flow transitions into an inertia-dominated regime, characterised by formation and spreading of circulations. Polymer solutions XG1500 and XG5000 exhibit more and larger circulations than their Newtonian analogue (XG1500N) because shear-thinning lowers viscosity at high shear, amplifying inertial effects at high flow rates. These wakes drive local pressure losses through shear-layer dissipation (Fig.~\ref{fig:OH_p_u1e-2}), identified as the primary contributor to the network’s nonlinear pressure drop~\citep[][]{matthai2024}. This will be discussed in Section~\ref{ssec3.3}. Although our 2D steady RANS setup captures only stationary recirculation zones at fracture irregularities and intersections, the very high local Reynolds numbers ($\mathrm{Re} \geq 10^4$) indicate a vortex-dominated, likely turbulent regime (Fig.~\ref{fig:OH_Re_u1e-2}). Scale-resolving simulations (LES/DES/DNS) would reveal unsteady vortices~\citep[][]{bui2025influence}; nevertheless, both signatures delineate secondary-flow regions driven by the coupling of fluid shear-thinning and inertia.

Additionally, shear-thinning helps the flow to overcome the viscous resistance of the highly tortuous fluid pathways~\cite[cf., ][]{matthai2024}, enabling the fluid to distribute more uniformly across branches compared to the Newtonian case (Fig.~\ref{fig:OH_partitioning}d-f). As a result, the flow of non-Newtonian fluids exhibits smaller velocities in individual active fractures. This finding highlights that shear-thinning can engender distributed network flow, manifesting a broad flow connectivity within the fracture system.

\begin{figure} 
	\centering
	\includegraphics[width=\columnwidth]{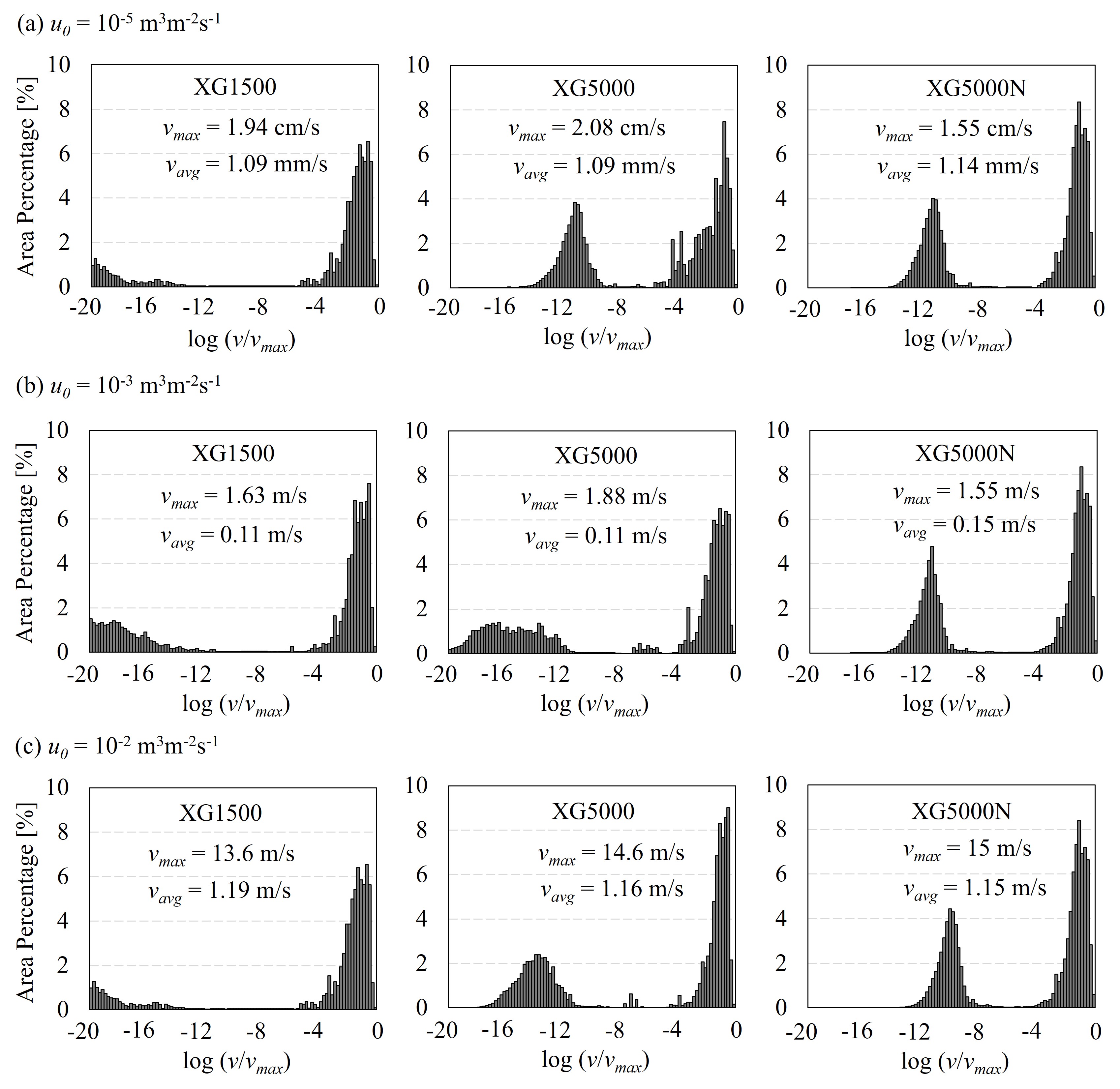}
	\caption{Hornelen network model: Velocity histograms for different fluids and influxes - (a) $u_0 = 10^{-5}$, (b) $10^{-3}$, and (c) $10^{-2}~\mathrm{m^3 m^{-2} s^{-1}}$. $v_{max}$ and $v_{avg}$ denote to the maximum and average velocities in the network, respectively. The velocity spectrum of the non-Newtonian fluids has more modes.}
	\label{fig:OH_vhistogram}
\end{figure}

For the Hornelen pattern, Non-Newtonian polymer solutions such as XG1500 and XG5000 give rise to a multimodal flow velocity distribution, reflecting the coexistence of unyielded and high-shear fast-flowing fractures, for a wide range of network influxes (Fig.~\ref{fig:OH_vhistogram}). The multiple high-velocity peaks mark a significant variability of fracture flow velocity, demonstrating that non-Newtonian fluids adapt more dynamically to complex fracture geometries than Newtonian ones, which retain a bimodal behaviour across all influxes. 


\subsubsection{\label{ssec3.3}Pressure drop and fracture flow regimes}

\begin{figure} 
	\centering
	\includegraphics[width=\columnwidth]{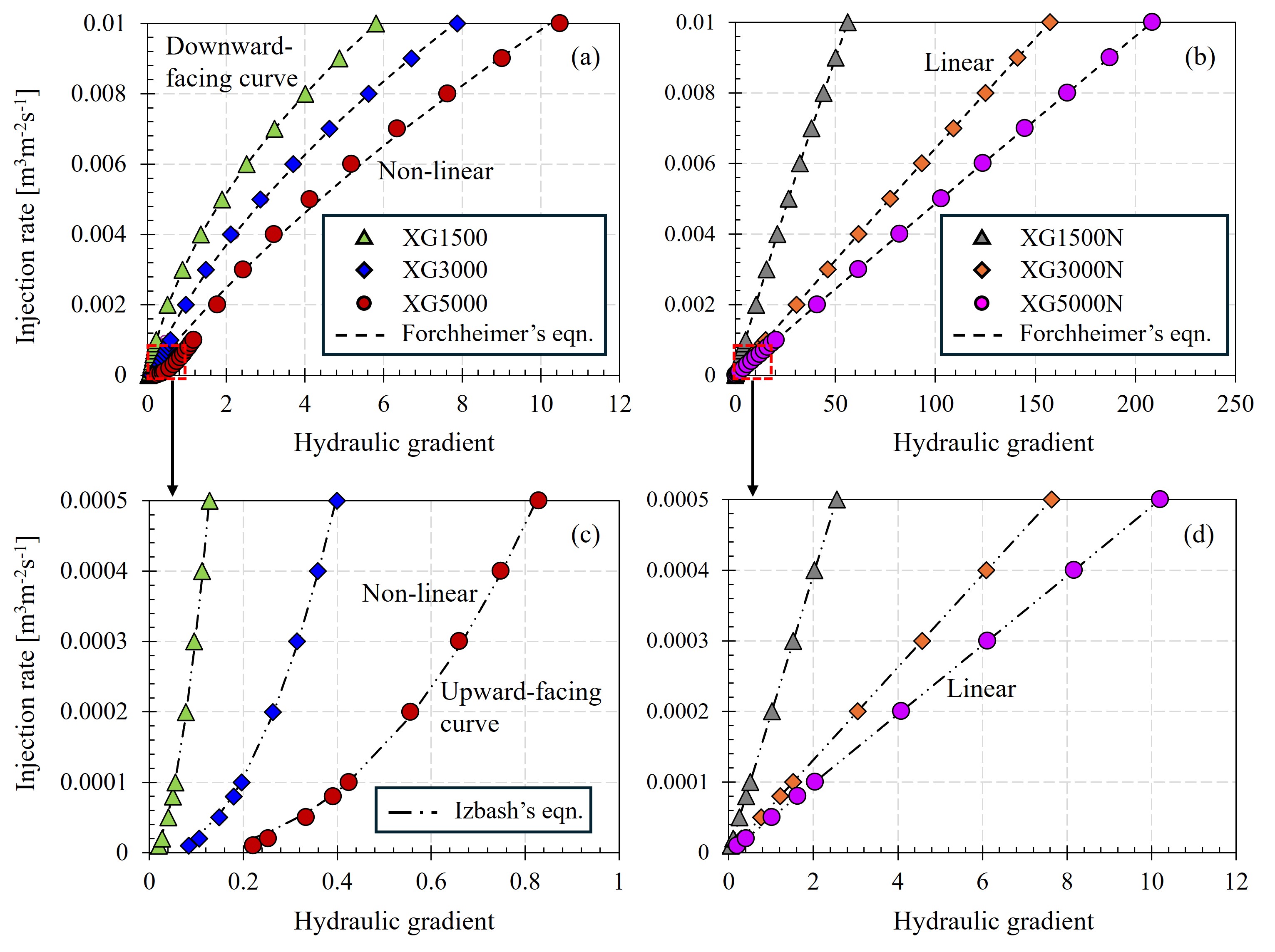}
	\caption{Hornelen network model: Correlation between the hydraulic gradient $J$ and injection rate $u_0$ for (a) non-Newtonian and (b) Newtonian fluids. (c) and (d) show magnified plots at low injection rates for (a) and (b), respectively.}
	\label{fig:OH_J}
\end{figure}

The relationship between network pressure drop and flow rate depends on both fluid rheology and rate itself, showing the nonlinearity and deviation from Darcy flow for non-Newtonian polymers but remaining linear for Newtonian fluids (Fig.~\ref{fig:OH_J}). This nonlinearity fingerprints the coexistance of different regimes in the network, arising from the combined effects of yield stress and shear-thinning rheology at different fracture flow velocities.

At $u_0 \leq 5 \times 10^{-4}~\mathrm{m^3 m^{-2} s^{-1}}$, the upward-facing curve (Fig.~\ref{fig:OH_J}c) for non-Newtonian polymer flows can be described by Izbash's formula~\citep[][]{izbash1931filtration}:

\begin{equation}
	u = \alpha~J^q \, ,
\end{equation}

where $\alpha$ is inversely correlated with network permeability~\citep[][]{siddiqui2016pre}. Parameter $q > 1$ (Table~\ref{tab:JF}) indicates that the overall flow resides in the pre-Darcy regime~\citep[][]{farmani2018analysis}. For a Newtonian fluid, this behaviour is observed only at extremely low rates, where localised phenomena such as pore compaction further hinder the flow~\citep[][]{kutilek1972non}. In flow of polymer solutions across static fractures, these viscous resistances are attributable to yield-stress-related features, which create high-viscosity and rigid regions, locally reducing fracture flowability.

\begin{table}[h!]
	\centering
	\begin{tabular}{l@{\hspace{1.5cm}}c@{\hspace{0.3cm}}c c}
		\hline
		\hline
		\multicolumn{3}{l}{\textbf{Low influxes :  $u_0 = \alpha J^q $ (Izbash's equation)}} & \\
		\hline
		Fluid & $\alpha$ & $q$ & $\mathrm{R^2}$\\
		\hline
		XG1500 & 2.50E-02 &  1.9 & 0.998 \\
		XG3000 & 4.02E-03 &  2.26 & 0.999 \\
		XG5000 & 7.99E-04 & 2.42 & 0.998 \\
		XG1500N & 1.97E-04 & 1 & 1 \\
		XG3000N & 6.56E-05 & 1 & 1 \\
		XG5000N & 4.90E-05 & 1 & 1 \\
		\hline 
		\hline
		\multicolumn{3}{l}{\textbf{High influxes:  $J = A u_0 + B {u_0}^2 $ (Forchheimer's equation)}}  & \\
		\hline
		Fluid & $A$ & $B$ & $\mathrm{R^2}$ \\
		\hline
		XG1500 & 1.77E+02 &  4.50E+05 & 1 \\
		XG3000 & 3.96E+02 &  3.85E+05 & 0.998\\
		XG5000 & 7.27E+02 & 2.97E+05 & 0.99 \\
		XG1500N & 5.05E+03 & 5.75E+05 & 1 \\
		XG3000N & 1.52E+04 & 5.11E+05 & 1 \\
		XG5000N & 2.04E+04 & 4.90E+05 & 1 \\
		\hline
		\hline
	\end{tabular}
	\caption{\label{tab:JF}DFN model: Estimated characteristic coefficients for Izbash's formula ($u_0 \leq 5 \times 10^{-4}~\mathrm{m^3 m^{-2} s^{-1}}$) and Forchheimer's formula ($u_0 > 5 \times 10^{-4}~\mathrm{m^3 m^{-2} s^{-1}}$) for various fluids. At low rates, $q > 1$ indicates the pre-Darcy flow dominates the fracture network.}
\end{table}


\begin{figure} 
	\centering
	\includegraphics[width=0.75\columnwidth]{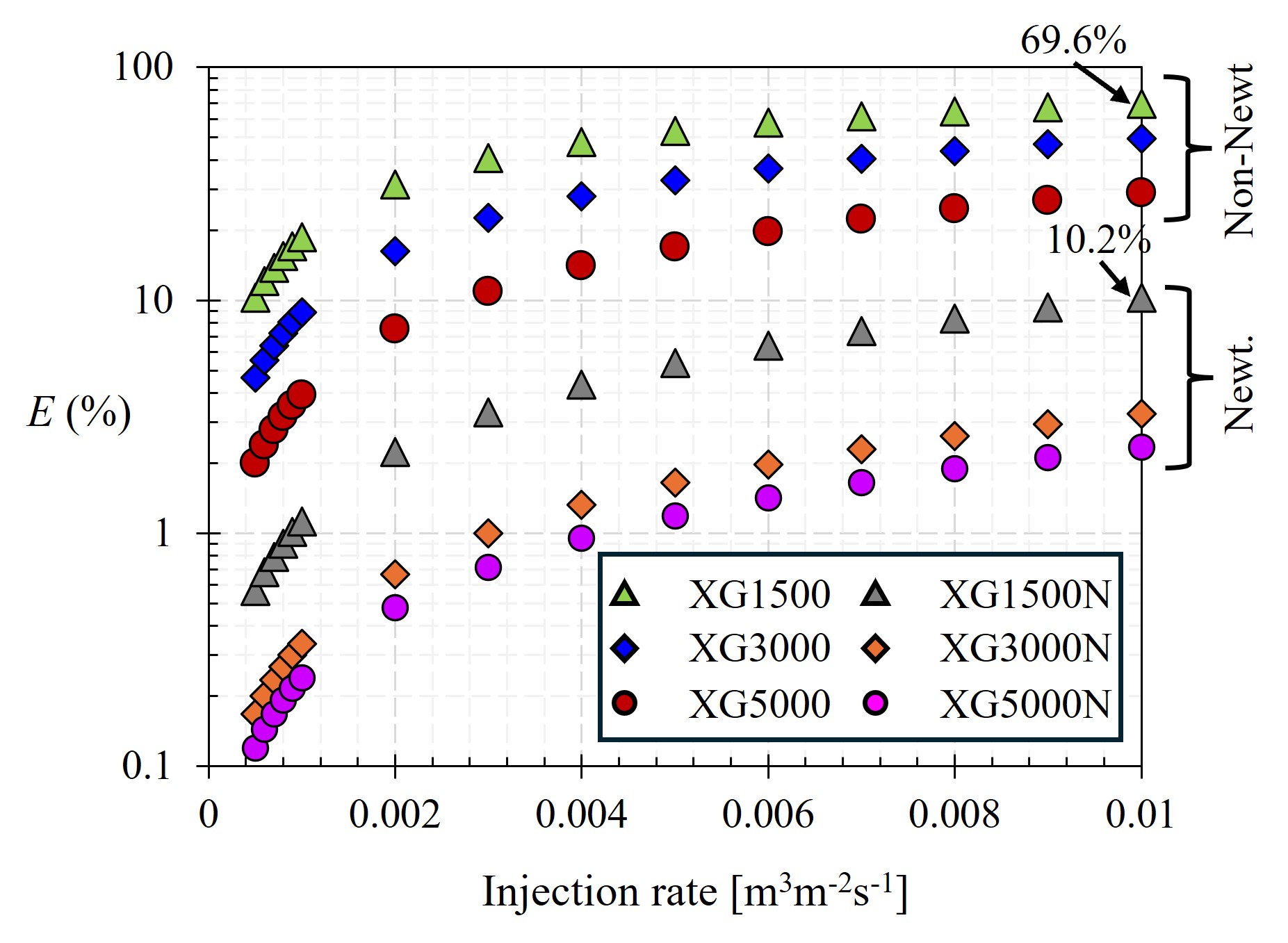}
	\caption{Hornelen network model: The contribution of inertia losses in total pressure drop, represented by the factor $E = \dfrac{B {u_0}^2 \times 100\%} {A {u_0} + B {u_0}^2} $ at $u_0 > 5 \times 10^{-4}~\mathrm{m^3 m^{-2} s^{-1}}$ for various fluids.}
	\label{fig:OH_E}
\end{figure}

At elevated flow rates, (i.e., $u_0 > 5 \times 10^{-4}~\mathrm{m^3 m^{-2} s^{-1}}$), the downward-facing curve (Fig.~\ref{fig:OH_J}a) fingerprints that inertial losses dominate the nonlinear relationship between pressure drop and flow rate~\citep[][]{matthai2024}. The non-Darcy (or post-Darcy~\citep[][]{kutilek1972non}) correlation between $J$ and $u_0$ can be described by Forchheimer's equation as~\citep[cf., ][]{whitaker1996forchheimer}:

\begin{equation}
	J = A u_0 + B {u_0}^2 \, ,
\end{equation}

with the linear term $A u_0$ representing the viscous losses, and the quadratic term $B {u_0}^2$ accounting for inertial effects. The coefficient $B$ characterises fracture flow nonlinearity, revealing the trend $B_\mathrm{XG1500} > B_\mathrm{XG3000} > B_\mathrm{XG5000}$, as the stronger shear-thinning behaviour and lower plastic viscosity of XG1500 reduce viscous forces and thus amplify fluid inertia effects (Table~\ref{tab:JF}). Among the three non-Newtonian fluids, inertial losses contribute the most to the total pressure drop for XG1500N, reaching $\sim 69.6\%$ at $u_0 = 10^{-2}~\mathrm{m^3 m^{-2} s^{-1}}$ (Fig.~\ref{fig:OH_E}). By contrast, due to their large constant viscosities, these losses account for only $ \leq 10.2 \%$ of the network pressure drop seen for Newtonian fluids.

Thus, fluid rheology is the primary factor driving the departure of network flow from Darcy's law as manifest in the nonlinear pressure-flow rate relationship across all influxes examined. However, the mechanisms underlying this nonlinearity differ between low and high flow rates.

\section{\label{sec:Section4}Discussion}

Recent fracture-flow studies such as \citep[][]{liu2016b,hyman2020flow,sherman2020characterizing,xue2022onset,matthai2024} only consider Newtonian fluids. The entirely new findings presented here highlight the significant impact of a more complex rheology on both local and network-wide flow  behaviours, addressing a major gap that practitioners working with non-Newtonian fluids live with. For the fracture pattern studied here, geometric complexity leads to a much wider spectrum of fluid deformation rates~\citep[cf., ][]{koyama2006numerical}, amplifying rheological responses, as compared with single-fracture behaviour~\citep[e.g., ][]{yan2008flow,lavrov2013numerical,lavrov2013redirection,zhang2023displacement}.

As the first attempt to examine viscoplastic fracture flow, our simulations confirm that yield stress rheology creates immobile rigid zones not only at dead-ends but also within interconnected fractures~\cite[cf., ][]{lavrov2023}, blocking fluid distribution to sub-branches and altering overall network connectivity. In naturally fractured reservoirs, such blockages would impede hydrocarbon flow and reduce fluid mobility~\citep[cf., ][]{paul2007fluid,xu2021blocking}, thereby diminishing oil and gas production. Reservoir fluids~\citep[e.g., ][]{hou2012experimental,mendes2017yield} and particle and polymer suspensions~\citep[e.g., ][]{hu2015effects,osiptsov2017fluid} can become trapped and remain in dead regions due to their very large yield stress~\citep[cf., ][]{majidi2010quantitative}, further limiting recovery and the recovery factor. Therefore, fluid rheology must be considered carefully when designing oil recovery processes involving proppant suspensions, heavy oil, and polymer-viscosified fluids.

Single-fracture studies and single-intersection flow analyses indicate that shear-thinning induces flow channeling and localisation~\citep[][]{auradou2008enhancement,lavrov2013redirection,lenci2022monte} by concentrating momentum in the fracture centre~\cite{zhang20193d} and by generating inertia-driven secondary flows that confine the bulk flow. By contrast, our network-flow simulations demonstrate that shear-thinning promotes a broader flow distribution, increasing flow into branches. This different behaviour arises because fluid inertia effects are enhanced by shear-thinning, allowing the fluid to overcome viscous resistances in tortuous pathways, facilitating a better spread of fluids across the network~\citep[cf. ][]{matthai2024}. 

It is important to note that the fracture pattern governs the influence of fluid inertia on network flow partitioning~\citep[][]{matthai2024}. For example, networks dominated by long, straight fractures exhibit a different behaviour than the Hornelen pattern studied here, preferentially reinforcing shear-thinning driven flow confinement to these straight paths~\citep[cf., ][]{rossen1992single}. The multimodal velocity distributions seen for Newtonian fluids become even more prominent for non-Newtonian ones, adapting to stronger local shear rate variations. This effect may even be more prominent in fracture networks with significant aperture heterogeneity~\citep[cf., ][]{matthai2024}. In our study, the aperture in individual fracture segments was held constant since in a 2D fracture model any constriction would choke off the flow in the entire fracture. Neglecting fracture roughness omits several aperture-scale effects; for instance, roughness can promote local channelisation in shear-thinning fluids as wider gaps lower apparent viscosity, while asperities create sub-yield shelters~\citep[][]{lavrov2013numerical,lavrov2013redirection,lavrov2023}. In yield-stress fluids, unyielded zones can form on fracture surfaces, fragmenting otherwise continuous plugs~\citep[][]{lavrov2013non,lavrov2023,nguyen2025viscoplastic}. In ongoing 3D simulation research, we further investigate how variable aperture and fracture geometry affect flow distribution and rheological responses.

Our simulation results also indicate that the limitations of reduced-order fracture models become particularly serious for non-Newtonian fluids. Thus, the shear-rate field is averaged across the fracture width and assumed uniform along the fracture wall, neglecting spatial variability~\citep[e.g., ][]{matthai2004,tabrizinejadas2023coupled}. In reality, local shear rate varies greatly near walls, bends, intersections, and converging–diverging zones, where changes in flow velocity create strong gradients~\citep[cf., ][]{matthai2024}. Obviously, a single effective shear rate cannot capture this nonlinear and spatially shear-dependent behaviour. For yield-stress fluids, the formation of unyielded rigid zones cannot yet be resolved with recuded-order models. Consequently, yield surfaces, wall effects, and the transition between flowing and solid-like behaviour are absent, leading to physically unrealistic behaviour.

While this study provides completely new insights into the fracture-network flow of complex non-Newtonian fluids, several practical limitations must be acknowledged. Fracture fluid rheology will change with reservoir conditions, such as temperature, pressure, and salinity, often deviating from laboratory measurements~\citep[][]{acharya1987rheology,ghoumrassi2015characterisation,zamani2017computation,skauge2018polymer,rock2020role}. These transformations are not considered by our model, limiting its ability to represent real-world scenarios. Given the very large deformation rates that network flow exhibits (i.e., $\dot{\gamma} > 10^4~\mathrm{s^{-1}}$), mechanical degradation can occur in non-Newtonian polymers~\citep[cf., ][]{druetta2019influence,aasen2019experimental,rock2020role,ferreira2022mechanical}, leading to a permanent loss of fluid viscosity and diminished rheological performance. This strain-history dependence has yet to be addressed. Moreover, future work should incorporate more sophisticated rheological models capable of capturing complex responses, such as limiting viscosity at extreme shear rates to improve predictive capability.

Shear-thinning potentially gives rise to time-dependent multi-scale vortices~\citep[cf., ][]{singh2017influence,bailoor2019vortex,patel2022vortex}. These cannot be captured by the steady-state simulations presented here. In the current work, a linear eddy-viscosity model (SST $k-\omega$ model) within the RANS framework was employed. This formulation assumes that Reynolds stresses are proportional to the mean rate of strain through a scalar eddy viscosity, extending a Newtonian-like analogy to turbulence. For non-Newtonian fluids, the local effective viscosity varies spatially and is directly incorporated into the governing equations, ensuring that rheological effects are consistently represented~\cite[cf., ][]{pakhomov2021rans}. We also note that our analysis focuses on water-based polymer suspensions, for which no-slip boundary conditions apply. While RANS provides a practical and computationally efficient approach for capturing inertia-dominated flow in complex fracture networks~\citep[][]{bui2025influence}, future studies should employ three-dimensional, scale-resolving simulations such as LES~\citep[e.g., ][]{amani2023large} or fully resolved DNS~\citep[e.g., ][]{ertl2018towards} to characterise transient, multi-scale vortical structures. This will also enable the investigation of shear-history-dependent rheology such as thixotropy~\citep[][]{kim2020non}, coupling with non-Newtonian rheology more realistically. Higher-resolution meshes are needed for a more accurate prediction of unyielded zones~\citep[cf., ][]{hoang2021lid,bui2023viscoplastic}. The current representation of fracture width with five layers of finite elements is not fully converged.


\section{\label{sec:Section5}Conclusion}

Fracture-network flow simulations with non-Newtonian fluids have been performed on a natural pattern, incorporating various dead-end and fracture intersection configurations. A broad range of injection rates has been studied. The fluid rheology, characterised by yield stress and shear-thinning behaviour was modelled using the Herschel-Bulkley-Papanastasiou approach. Our study contributes to closing a major research gap in the understanding of non-Newtonian fluid flows in a fracture system and reveals that:

\begin{enumerate}
	\item Due to the network geometric complexity, shear rate varies richly across the fracture network, spanning many orders of magnitude. For non-Newtonian aqueous polymer solution, this variation creates significant viscosity contrasts, with high viscosities dominating low-shear regions and low viscosities occuring in localised shear zones;
	\item At low injection rates, the yield stress effect dominates fluid rheology, leading to the formation of rigid zones, occupying up to $\sim 65\%$ of the network area at mm/s velocity ranges. These zones can remain static at dead-ends and block interconnected fractures, significantly restricting network flow connectivity, or they can move with the fluid. As fracture flow velocities increase to m/s, the extent of these zones diminishes. Only in dead-ends, the rigid zones persist;
	\item At high injection rates, shear-thinning rheology becomes the dominant factor shaping flow behaviour by reducing viscous effects in high-shear regions. Across the network, fracture flow becomes inertia-dominated, forming secondary swirling flows around intersections;
	\item Depending on local flow rate, fluid rheology alters network flow partitioning through different mechanisms. At low rates, yield stress causes blockage in side branches, confining the flow to specific dominant pathways and reducing overall network connectivity. At high rates, shear-thinning rheology allows the flow to overcome viscous resistance and enhance fluid inertia, navigating tortuous pathways more effectively across the network;
	\item At the network scale, fluid rheology promotes a multimodal velocity distribution induced by the interplay between fluid properties and network geometry. The pressure drop-flow rate relationship remains nonlinear across all influxes, stemming from pre-Darcy behaviour due to yield stress at low rates and inertia amplified by shear-thinning at high rates.
\end{enumerate}

\section*{Acknowledgements}


This work is funded by the Australian Research Council under grant number DP220100851. CMB wishes to acknowledge the support provided by the Melbourne Research Scholarship and Research Computing Services at The University of Melbourne. Contour visualisation was performed using Paraview (paraview.org).

\bibliography{mybibfile}

%

\end{document}